\pdfoutput=1

\documentclass[preprint]{elsarticle}

\biboptions{round, semicolon, authoryear}   

\usepackage{amsmath}
\usepackage{amsfonts}
\usepackage{amssymb}

\usepackage[T1]{fontenc}

\usepackage{color, soul}

\usepackage{url}    

\usepackage[margin=1in]{geometry}    

\usepackage{graphicx}    

\usepackage{booktabs}    
\usepackage{caption}

\usepackage{array}
\newcolumntype{K}[1]{>{\raggedright\let\newline\\\arraybackslash\hspace{0pt}}m{#1}}
\newcolumntype{Q}{>{\centering\arraybackslash}X}

\usepackage{tabularx}    
\usepackage{tabulary}     

\usepackage{enumitem}
\newenvironment{QandA}{\begin{enumerate}[label=\bfseries\alph*.]\bfseries}
                      {\end{enumerate}}
\newenvironment{answered}{\par\normalfont}{}

\usepackage[toc, page]{appendix}

\begin{document}

\begin{frontmatter}

\title{Check yourself before you wreck yourself: \\Assessing discrete choice models through predictive simulations}

\author[tbrathwaite]{Timothy Brathwaite\corref{cor1}}
\ead{timothyb0912@gmail.com}

\cortext[cor1]{Corresponding Author}

\address[tbrathwaite]{Lyft Inc.\\ 185 Berry Street, Suite 5000, San Francisco, CA, 94107}

\begin{abstract}
Typically, discrete choice modelers develop ever-more advanced models and estimation methods. Compared to the impressive progress in model development and estimation, model-checking techniques have lagged behind. Often, choice modelers use only crude methods to assess how well an estimated model represents reality. Such methods usually stop at checking parameter signs, model elasticities, and ratios of model coefficients. In this paper, I greatly expand the discrete choice modelers' assessment toolkit by introducing model checking procedures based on graphical displays of predictive simulations. 

Overall, my contributions are as follows. Methodologically, I introduce a general and `semi-automatic' algorithm for checking discrete choice models via predictive simulations. By combining new graphical displays with existing plots, I introduce methods for checking one's data against one's model in terms of the model's predicted distributions of choices ($P \left( Y \right)$), choices given explanatory variables ($P \left( Y \mid X \right)$), and explanatory variables given choices ($P \left( X \mid Y \right)$). Empirically, I demonstrate my proposed methods by checking the models from \citet{brownstone_forecasting_1998}. Through this case study, I show that my proposed methods can point out lack-of-model-fit in one's models and suggest concrete model improvements that substantively change the results of one's policy analysis. Moreover, the case study highlights a practical trade-off between precision and robustness in model checking.
\end{abstract}

\end{frontmatter}

\section{Introduction}
\label{sec:intro}
``Check. Your. Work. It is okay to make mistakes, but it is not okay to turn in work that is riddled with mistakes that checking your work could have caught.'' These sage words were spoken by my undergraduate mathematics professor, Dr. Kenneth Holladay. Unfortunately, somewhere between Dr. Holladay's calculus classes and my graduate discrete choice studies, checking one's work was demoted from a main actor to a supporting or missing role in the ever-unfolding data analysis drama.

The tragedy of this situation is that regardless of whether one is working on undergraduate calculus problems or building predictive models for governments and corporations, there is a need to check one's work. From an econometric perspective, we do not typically have economic theory to guide all aspects of our model specification \citep[p.85]{dagsvik_2017_invariance}. Accordingly, it would be wise to check the implications of our (likely) mis-specified models. And from a broader statistical view, it is often noted that, ``all models [including discrete choice models] are false, but some are useful'' \citep{box_science_1976}. Again, we find checking for \textit{correctness} or \textit{truth} of one's model to be hopeless a priori, and the best we can do is check how our models misrepresent reality \citep[pp.734, 800]{gelman_posterior_1996}.

\begin{figure}
\centering
\includegraphics[width=0.75\textwidth]{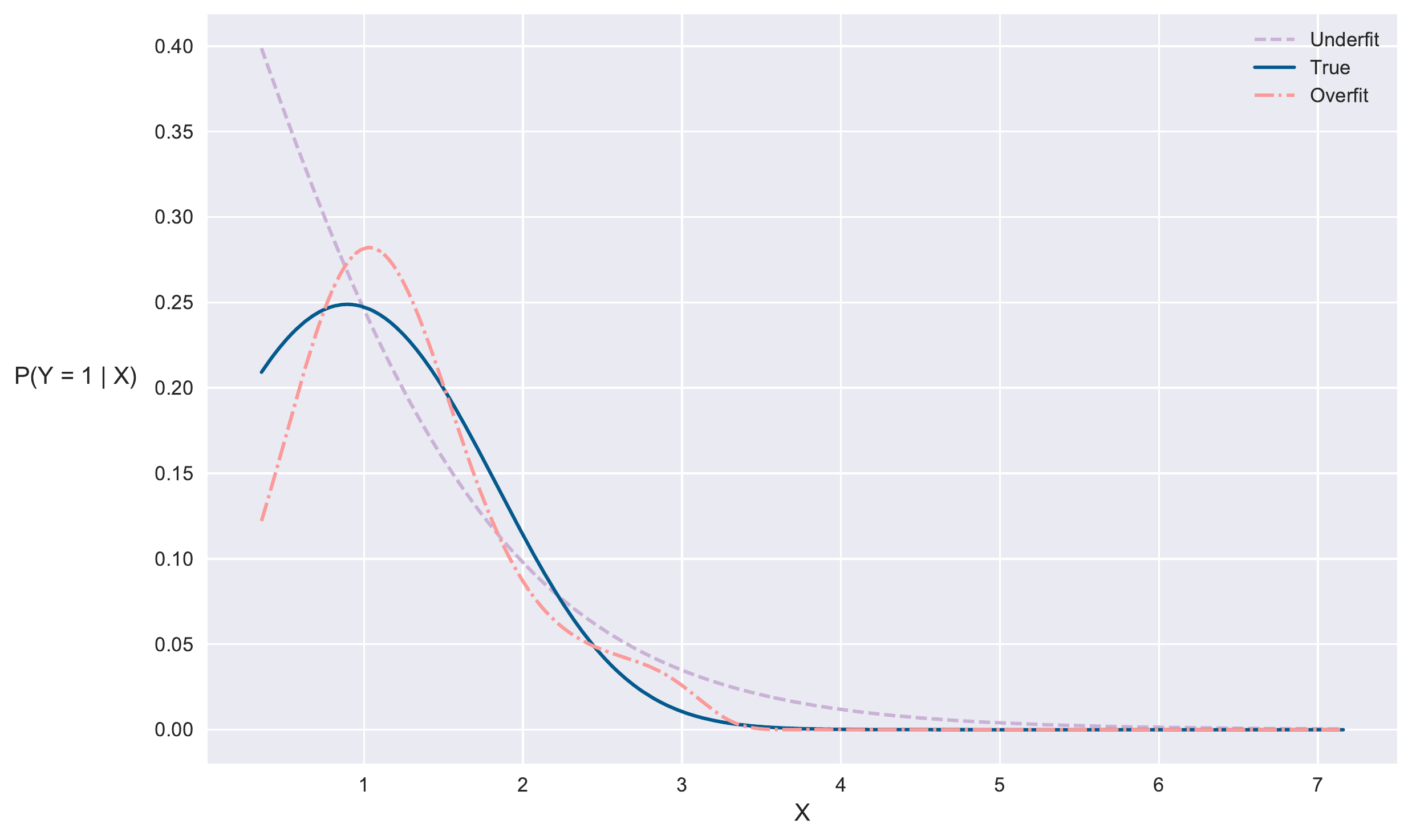}
\caption{Example of underfitting and overrfitting}
\label{fig:underfitting}
\end{figure}

For discrete choice models, many such checks can be broadly categorized as checking for ``overfitting'' or ``underfitting.'' overfitting is the degree to which one's model performs better when evaluated on data used for estimation (i.e. training or in-sample data) as compared to data that was not used for estimation (i.e. testing or out-of-sample data) \citep[p.6]{fariss_enhancing_2017}. Colloquially, severe overfitting is characteristic of models that ``learned'' the patterns in the in-sample data at the expense of patterns that generalize to out-of-sample data. Conversely, underfitting is the degree to which one's model has failed to learn patterns in one's in-sample data \citep[p.6]{fariss_enhancing_2017}. Underfit models typically perform poorly on some problem-specific metric for both in-sample and out-of-sample data. See Figure \ref{fig:underfitting} for an illustrative depiction of these two concepts in a binary discrete choice setting. Here, the data generating relationship is shown in dark blue. The underfit model, shown in dashed purple, is overly simple and has failed to capture observed patters in the data. Conversely, the overfit model is overly complex; in dashed and dotted light orange, the overfit model captures spurious patterns in the data that will not generalize to new samples from the data generating process.

Currently, numerous methods exist for diagnosing overfitting. Such techniques include cross-validation \citep{stone_cross_1974}, resampling-methods such as the ``0.632+'' bootstrap \citep{efron_improvements_1997}, and the use of held-out data \citep{steckel_cross_1993}. Importantly, these techniques are highly general. Instead of testing for specific, hypothesized causes of overfitting, these methods check for overfitting's presence and extent. This generality promotes use: researchers can use the methods regardless of the specific analyses\footnote{Of course, one's analytic task can impact the way such checks are implemented. For instance, cross-validation is performed differently when dealing with cross-sectional panel data as compared to purely cross-sectional data.} being performed. As a result, the overall quality of predictive analyses has increased. A mostly standardized tool-box and language for assessing overfitting has developed, enabling researchers to better evaluate predictive models.

Comparatively, checks for underfitting are far less common. For instance, within the discrete choice literature, few checks for the general presence and extent of underfitting have been proposed. As examples, consider two popular discrete choice textbooks: ``Discrete Choice Methods with Simulation, 2nd Edition'' by Kenneth Train \citeyearpar{train_discrete_2009} and ``Discrete Choice Analysis: Theory and Application to Travel Demand'' by \citet*{ben-akiva_discrete_1985}. In \citet{train_discrete_2009}, only 1 of the 90 sections in the book (Section 11.5) discusses model checking in a manner that can be described as a general check for underfitting. Even then, the checks discussed in that section are only applicable to a particular class of models---mixed logit models. Now consider \citet{ben-akiva_discrete_1985}. In that text, only 1 of 78 sections (Section 7.6) describes procedures for directly checking one's model against one's data for lack-of-fit. Unfortunately, this paucity of methods for assessing underfitting goes beyond these two textbooks. It is characteristic of the field.

To improve the state of practice, I propose and demonstrate methods for using predictive simulations to perform general checks of underfitting in discrete choice models. In particular, I make the following contributions.  Methodologically, I introduce a general and `semi-automatic' algorithm for checking discrete choice models via predictive simulations. Moreover, by combining new graphical displays with existing techniques, I introduce methods for checking one's data against one's model in terms of the model's predicted distributions of choices ($P \left( Y \right)$), choices given explanatory variables ($P \left( Y \mid X \right)$), and explanatory variables given choices ($P \left( X \mid Y \right)$). Empirically, I demonstrate my proposed methods by checking the models from \citet{brownstone_forecasting_1998}. Through this case study, I show that my proposed methods can point out lack-of-model-fit in one's models and suggest concrete model improvements that substantively change the results of one's policy analysis.

Importantly, these model checking techniques complement rather than replace careful economic and behavioral scrutiny of one's models. By ensuring adherence with sound economic principles while also diagnosing and reducing the amount of underfitting in our models, we increase our chances of creating maximally useful representations of reality.

The remainder of this paper is structured as follows. To ease discussion of this paper's ideas, Section \ref{sec:methods} immediately describes the proposed methodology and provides examples of the graphical displays introduced in this paper. Once readers are familiar with the proposed simulation-based checks for underfitting, Section \ref{sec:lit-review} relates these methods to existing model tests and assessments in the discrete choice and statistics literature. Next, Section \ref{sec:case-study-part2} will continue this paper's case study, showing how the displays from Section \ref{sec:methods} suggest avenues for model-improvement, not merely model criticism. Section \ref{sec:conclusion} will then recap the paper's contributions and conclude.

Finally, Appendix \ref{sec:freq-questions} contains answers to frequently asked questions about the interpretation and use of predictive simulations for model checking, and Appendix \ref{sec:appendix-b-expanded-results} contains model estimation results that are of secondary importance to the paper but perhaps of interest to some readers.

\section{Methodology: Demonstration and Description}
\label{sec:methods}
In this section, I demonstrate and describe this paper's proposed methodology for assessing underfitting in discrete choice models. It cannot be reiterated strongly enough that the \textbf{\textit{main focus}} of this paper's techniques is to show and quantify the existence and extent of underfitting of one's model. As a result, this section will not contain any information about how one might use the proposed techniques to help suggest model improvements. Such a discussion is of secondary importance to this paper and is deferred to Section \ref{sec:case-study-part2}. Moreover, not all techniques in this section will provide obvious clues to how one can improve one's model. This is to be expected. As an analogy, consider thermometers. They are useful for detecting the presence and severity of a fever, even though they don't suggest strategies for curing the underlying problem. Likewise, this paper's methods are valuable for detecting and measuring underfitting, regardless of the degree to which they suggest ways to fix the underlying causes of the lack-of-fit.

Disclaimers aside, I use a case study of vehicle choice by \citet{brownstone_forecasting_1998} to provide an initial illustration of the proposed techniques. Though published twenty years ago, I chose \citeauthor{brownstone_forecasting_1998}'s paper for three reasons: (1) because the data is freely available for download by the public, (2) because the article is well-known and well-cited within the discrete choice community, and (3) because the article represents standard best practices within the field. As such, by applying this paper's proposed model checking techniques to the vehicle choice models of \citet{brownstone_forecasting_1998}, we gain insight into the types of underfitting that one might find, even in models built by the best of us.

Accordingly, Subsection \ref{sec:case1-part1-data} describes the vehicle choice data and model from \citet{brownstone_forecasting_1998}. Next, Subsections \ref{sec:case1-part1-lp-plots} - \ref{sec:case1-part1-cdfs} show the seven types of plots that will be used in this paper to assess a model's lack-of-fit. These plots are introduced in increasing order of granularity. For example, the first two plots assess whether one's model fits one's data in an aggregate sense; the second and third plots assess a model's lack-of-fit for particular groups of observations for a given alternative; and the final three plots assess a model's lack-of-fit with respect to a particular alternative and particular values of a chosen explanatory variable.

Finally, once specific examples of the various plots have been introduced, Subsection \ref{sec:case1-part1-discussion} will discuss the results and interpretations of these plots for diagnosing the character and extent of underfitting in \citet{brownstone_forecasting_1998}'s multinomial logit (MNL) model. Subsection \ref{sec:general-methodology} will then abstract from the case study to present this paper's general methodology for checking underfitting in discrete choice models. To make the methods as widely useful as possible, special attention will be paid to how the general method may be adapted for use with `unlabelled alternatives.' Moreover, Subsection \ref{sec:general-methodology} will present a semi-automatic algorithm for applying this paper's methodology, further decreasing the minimum efforts needed to apply this paper's techniques.

\subsection{A Case Study of \citet{brownstone_forecasting_1998} (Part 1)}
\label{sec:case1-part1}

\subsubsection{Data and modeling context}
\label{sec:case1-part1-data}
The data in this case study comes from a stated-preference study of household attitudes and preferences for alternative-fuel vehicles \citep{brownstone_forecasting_1998, brownstone_transactions_1996}. The sampling plan used random telephone-digit dialing followed by a postal survey to collect data from 4,654 households. The collected data included each household's choice amongst six alternatives (three different fuel types and two body types per fuel type), the attributes of each alternative, and some limited characteristics of the household. For a detailed description of the collected attributes and characteristics, see Table \ref{table:variable-definitions-vehicle-choice}. Together, \citeauthor{brownstone_forecasting_1998} used the collected\footnote{For further information about the data and collection protocol, see \citet{brownstone_forecasting_1998, brownstone_transactions_1996}. To access the raw data, visit the website of the Journal of Applied Econometrics and download the data associated with \citet{mcfadden_mixed_2000}.} data to estimate four vehicle choice models: a MNL model, a probit model, and two mixed logit models.

\begin{table}
\centering
\begin{tabular}{l K{0.75\linewidth}}
\toprule
{Variable} & {Definition}\\
\midrule

Price/log(income) & Purchase price (in thousands of dollars) divided by log(household income in thousands) \\
Range & Hundreds of miles vehicle can travel between refuellings/rechargings \\
Acceleration & Tens of seconds required to reach 30 mph from stop \\
Top speed & Highest attainable speed in hundreds of MPH\\
Pollution & Tailpipe emissions as fraction of those for new gas vehicle \\
Size & 0=mini, 0.1 =subcompact, 0.2=compact, 0.3=mid-size or large \\
`Big enough' & 1 if household size is over 2 and vehicle size is 3; 0 otherwise \\
Luggage space & Luggage space as fraction of comparable new gas vehicle \\
Operating cost & Cost per mile of travel (tens of cents): home recharging for electric vehicle; station refuelling otherwise \\
Station availability & Fraction of stations that can refuel/recharge vehicle\\
Sports utility vehicle & 1 if sports utility vehicle, 0 otherwise\\
Sports car & 1 if sports car\\
Station wagon & 1 if station wagon\\
Truck & 1 if truck\\
Van & 1 if van\\
EV & 1 if electric vehicle (EV)\\
Commute $< 5$ \& EV & 1 if electric vehicle and commute $< 5$ miles/day\\
College \& EV & 1 if electric vehicle and some college education\\
CNG & 1 if compressed natural gas (CNG) vehicle\\
Methanol & 1 if methanol vehicle\\
College \& Methanol & 1 if methanol vehicle and some college education\\

\bottomrule
\end{tabular}
\caption{Variable Definitions}
\label{table:variable-definitions-vehicle-choice}
\end{table}

For part one of this case study, I demonstrate this paper's methods with \citeauthor{brownstone_forecasting_1998}'s MNL model\footnote{However, I return to the assessment of \citeauthor{brownstone_forecasting_1998}'s chosen mixed logit model in Section \ref{sec:case-study-part2}.}. I use this model because it is likely to be familiar to the largest spectrum of readers. Moreover, producing the predictive simulations for \citeauthor{brownstone_forecasting_1998}'s chosen mixed logit model seemed unnecessarily tedious. This would require writing custom code to estimate their model, yet the MNL can be used just as effectively to demonstrate this paper's model checking techniques. 

Justifications aside, Table \ref{table:vehicle-choice-mnl-results}, shows the results of reproducing \citeauthor{brownstone_forecasting_1998}'s MNL model. As expected, the results match those of the original study. Moreover, almost all variables in the model are statistically significant, and no variables have an obviously wrong sign. \citeauthor{brownstone_forecasting_1998}'s MNL model therefore seems to be a good `working model' that one can begin refining by iteratively: (1) criticizing the working model, (2) positing and estimating a new model that addresses the earlier criticisms, and (3) repeating steps 1 and 2 \citep[p.795]{box_science_1976}.

\begin{table}
\centering
\begin{tabular}{lrc}
\toprule
{Variable} &  Estimate &  Std. err \\
\midrule
Price over log(income)       &      -0.185** &    0.027 \\
Range (units: 100mi)         &       0.350** &    0.027 \\
Acceleration (units: 0.1sec) &      -0.716** &    0.111 \\
Top speed (units: 0.01mph)   &       0.261** &    0.081 \\
Pollution                    &      -0.444** &    0.102 \\
Size                         &       0.934** &    0.316 \\
Big enough                   &       0.143\hphantom{**} &    0.077 \\
Luggage space                &       0.501** &    0.191 \\
Operation cost               &      -0.768** &    0.076 \\
Station availability         &       0.413** &    0.096 \\
Sports utility vehicle       &       0.820** &    0.141 \\
Sports car                   &       0.637** &    0.148 \\
Station wagon                &      -1.437** &    0.062 \\
Truck                        &      -1.017** &    0.049 \\
Van                          &      -0.799** &    0.047 \\
EV                           &      -0.179\hphantom{**} &    0.172 \\
Commute < 5 \& EV             &       0.198*\hphantom{*} &    0.084 \\
College \& EV                 &       0.443** &    0.109 \\
CNG                          &       0.345** &    0.092 \\
Methanol                     &       0.313** &    0.103 \\
College \& Methanol           &       0.228*\hphantom{*} &    0.089 \\

\multicolumn{3}{c}{}\tabularnewline
Log-likelihood & -7,391.830\hphantom{**} & {} \tabularnewline

\bottomrule
\multicolumn{3}{l}{Note: * means $\textrm{p-value} < 0.05$ and ** means $\textrm{p-value} < 0.01$.} \\
\end{tabular}
\caption{\citeauthor{brownstone_forecasting_1998}'s MNL model}
\label{table:vehicle-choice-mnl-results}
\end{table}

Before seeing any data, \citeauthor{brownstone_forecasting_1998} were critical of the MNL model because its ``independence from irrelevant alternatives'' (IIA) property implies restrictive substitution patterns among the various alternatives being modeled \citep[pp.109-110]{brownstone_forecasting_1998}. This a-priori criticism led \citeauthor{brownstone_forecasting_1998} to estimate the probit and mixed logit models mentioned above since those models do not have the IIA property. To complement such data-agnostic reasons for model refinement, this paper suggests a data-driven, graphical approach to model criticism by using predictive simulations to check for underfitting. In Subsections \ref{sec:case1-part1-lp-plots} - \ref{sec:case1-part1-cdfs} that follow, I will use  \citeauthor{brownstone_forecasting_1998}'s MNL model to showcase seven graphical model-checks that I believe can be widely useful for discrete choice models. Then, Subsection \ref{sec:general-methodology} will generalize these plots by detailing an overarching methodology for model checking via predictive simulations.

\subsubsection{Log-predictive plots}
\label{sec:case1-part1-lp-plots}
This paper's first type of graphical model check is the log-predictive plot. This plot compares a scalar summary of the predictive ability of a given model on the observed data against the distribution formed by evaluating the scalar measure on simulated datasets. Because this plot evaluates a single number, the observed log-predictive value, this plot is amongst the most aggregate checks of underfitting.

\begin{figure}
\centering
\includegraphics[width=0.75\textwidth]{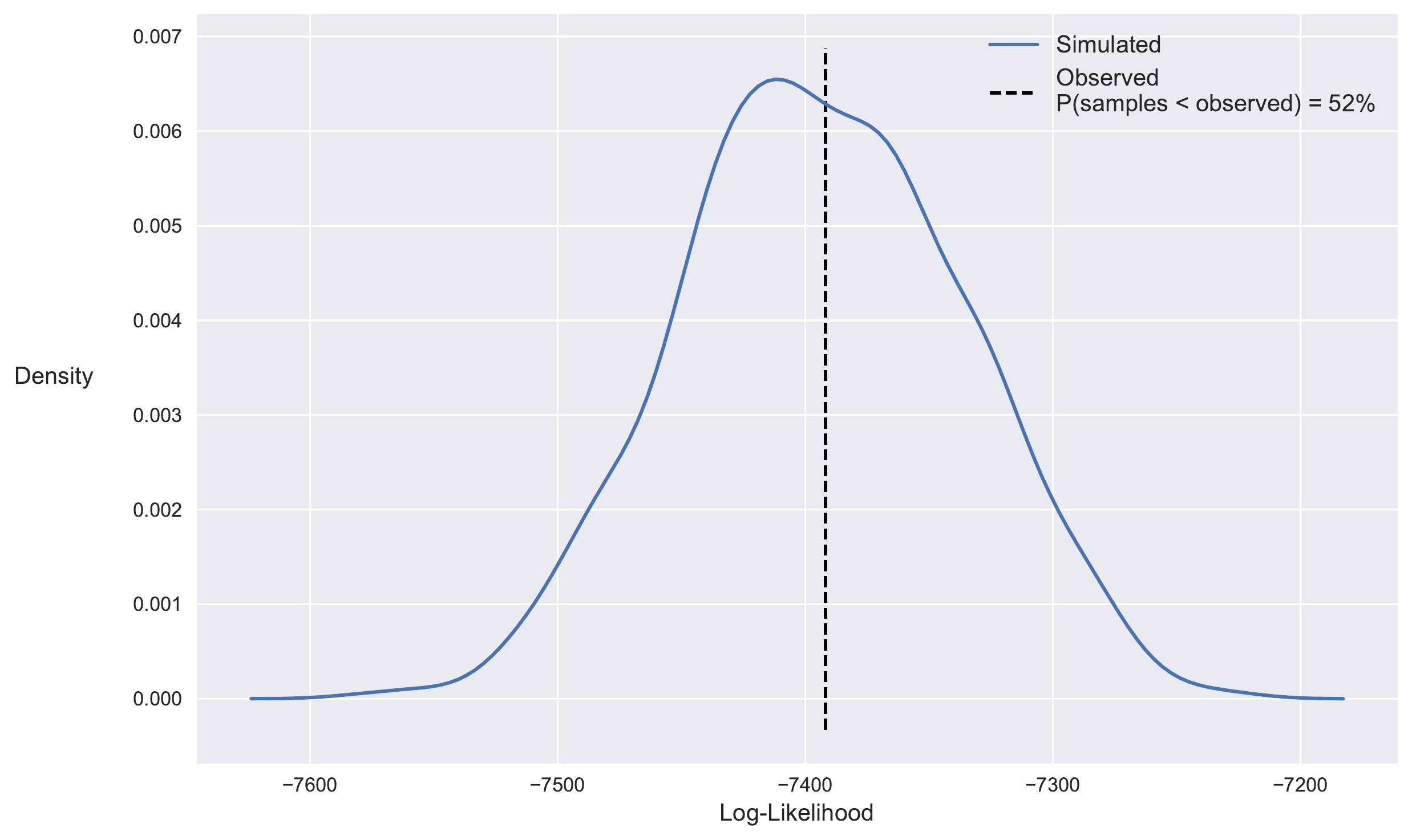}
\caption{Log-predictive plot of \citeauthor{brownstone_forecasting_1998}'s MNL model}
\label{fig:orig-mnl-log-predictive}
\end{figure}

For a concrete example, consider Figure \ref{fig:orig-mnl-log-predictive}. Here, the scalar measure is the log-likelihood, $T \left( X, Y; \beta _{\textrm{MLE}} \right) = \sum _i \sum _{j \in C_i} y_{ij} \ln \left[ P \left( y_{ij} = 1 \mid X_{ij}, \beta _{\textrm{MLE}} \right) \right]$. In this equation, $X$ is the design matrix for the observed data; $Y$ is a vector of outcomes with one element per observation ($i$) per available alternative ($j$); $y_{ij} \in \left\lbrace 0, 1 \right\rbrace$ denotes whether observation $i$ chose alternative $j$; $X_{ij}$ contains the explanatory variables for observation $i$ and alternative $j$; $C_i$ is the choice set for individual $i$; $\beta _{\textrm{MLE}}$ is the maximum likelihood estimate (MLE) displayed in Table \ref{table:vehicle-choice-mnl-results}; and $P \left( y_{ij} = 1 \mid X_{ij}, \beta _{\textrm{MLE}} \right)$ is calculated from \citeauthor{brownstone_forecasting_1998}'s MNL model.

The observed log-likelihood, $T \left( X, Y; \beta _{\textrm{MLE}} \right)$, is compared against the distribution formed by computing $T \left( X, Y^r; \beta _{\textrm{MLE}} \right)$ for simulated outcomes $Y^r$, for draws $r \in \left\lbrace 1, 2, ..., R \right\rbrace$. Here, $R$ is the total number of simulated datasets. In this case, we first simulate values of the choice model parameters ($\beta$) from the asymptotic sampling distribution of $\beta _{\textrm{MLE}}$. In other words, if $\mathcal{MVN} \left( \mu, \Sigma \right)$ is a multivariate normal distribution with mean $\mu$ and covariance matrix $\Sigma$, then I draw $\beta ^r$ from $\mathcal{MVN} \left(  \beta _{\textrm{MLE}}, -H^{-1} \right)$ where $-H^{-1}$ is the negative inverse of the Hessian matrix, evaluated at $\beta _{\textrm{MLE}}$. Then for each draw of $\beta ^r$, I compute $P \left( y_{ij} = 1 \mid X_{ij}, \beta ^r \right) \ \forall \  i, j \ \in C_i$. Finally, for each observation $i$, I simulate an outcome $y ^r_i = \left[ y ^r _{i1}, y ^r _{i2}, ..., y ^r _{i, \mid C_i \mid} \right]$ from the categorical distribution with probabilities $P \left( y_{ij} = 1 \mid X_{ij}, \beta ^r \right) \  \forall j \ \in C_i$. Once the simulated values of $T \left( X, Y^r; \beta _{\textrm{MLE}} \right)$ have been calculated, their distribution can be displayed using histograms, kernel density estimates, and/or empirical cumulative density functions. In Figure \ref{fig:orig-mnl-log-predictive}, a kernel density estimate was used.

Mechanics aside, the rationale behind the log-predictive plots is as follows. If one's model fits one's data well, then the simulated outcomes $\left\lbrace Y^r \right\rbrace$ from one's model should be similar to one's observed outcomes. The log-predictive plots measure similarity in terms of some log-predictive measure such as the log-likelihood evaluated at the MLE of the observed data. In a bayesian context, one might instead use the log pointwise-predictive density $T \left( X, Y \right) = \sum _i \sum _{j \in C_i} y_{ij} \ln \left[ \frac{1}{S }\sum _{s=1} ^S P \left( y_{ij} = 1 \mid X_{ij}, \beta _s \right) \right]$ where $S$ is the total number of parameter samples from one's posterior/prior distribution and $\beta _s$ is a single sample \citep[p.169]{gelman_bayesian_2014}. In both bayesian and frequentist analyses, one expects that, for a well-fitting model, the distribution of log-predictive values formed by $T \left( X, Y^r \right)$ will be symmetric and unimodal with its mode at the observed value $T \left( X, Y \right)$. This would indicate that the log-predictive values of the observed outcomes are very likely, and that one's model is not biased towards producing outcomes that have consistently lower or higher log-predictive values. Deviations from this ideal represent model misfit.

As exemplified by Figure \ref{fig:orig-mnl-log-predictive}, checking an estimated model with a log-predictive plot does not always reveal a lack of fit\footnote{On the other hand, log-predictive plots \textbf\textit{{can}} reveal a lack of fit. The expectation of the distribution of simulated log-predictive values is not the observed log-predictive value. This is true despite the fact that the expectation of the simulated parameters is the point-estimate of the model parameters. As is well-known, $ E \left[ f \left( x \right) \right] \neq f \left( E \left[ x \right] \right)$. That is to say, the expectation of a function need not equal the function evaluated at an expectation}. This is because frequentist analyses typically optimize for the log-likelihood, and even bayesian analyses revise their prior distributions to concentrate in areas of the parameter space with higher log-likelihoods (and hence high log pointwise-predictive values). However, if one's log-predictive plot of an estimated model displays a marked lack of fit, this means that something is wrong. For example one's posited model may be woefully inadequate, one's estimation routine may not be converging, or one's computations may be suffering from numeric errors. In this sense, log-predictive plots are basic, graphical sanity-checks.

Beyond checking one's estimated models, log-predictive plots are most useful for checking one's prior distributions in a bayesian analysis. In this setting, where analysts often choose their prior distributions for convenience rather than to reflect their true beliefs, the log-predictive plots can reveal large discrepancies between one's implied prior beliefs about the outcomes for one's dataset and the actual outcomes. If one notices such a conflict between one's data and one's prior after (1) observing a lack-of-fit between one's posterior distribution and one's data and (2) ruling out a lack-of-fit between one's likelihood and one's data, then one's `convenience prior' should be adjusted to reflect one's true a-priori information and beliefs.

\subsubsection{Market share plots}
\label{sec:case1-part1-ms-plots}
\begin{figure}
\centering
\includegraphics[width=0.75\textwidth]{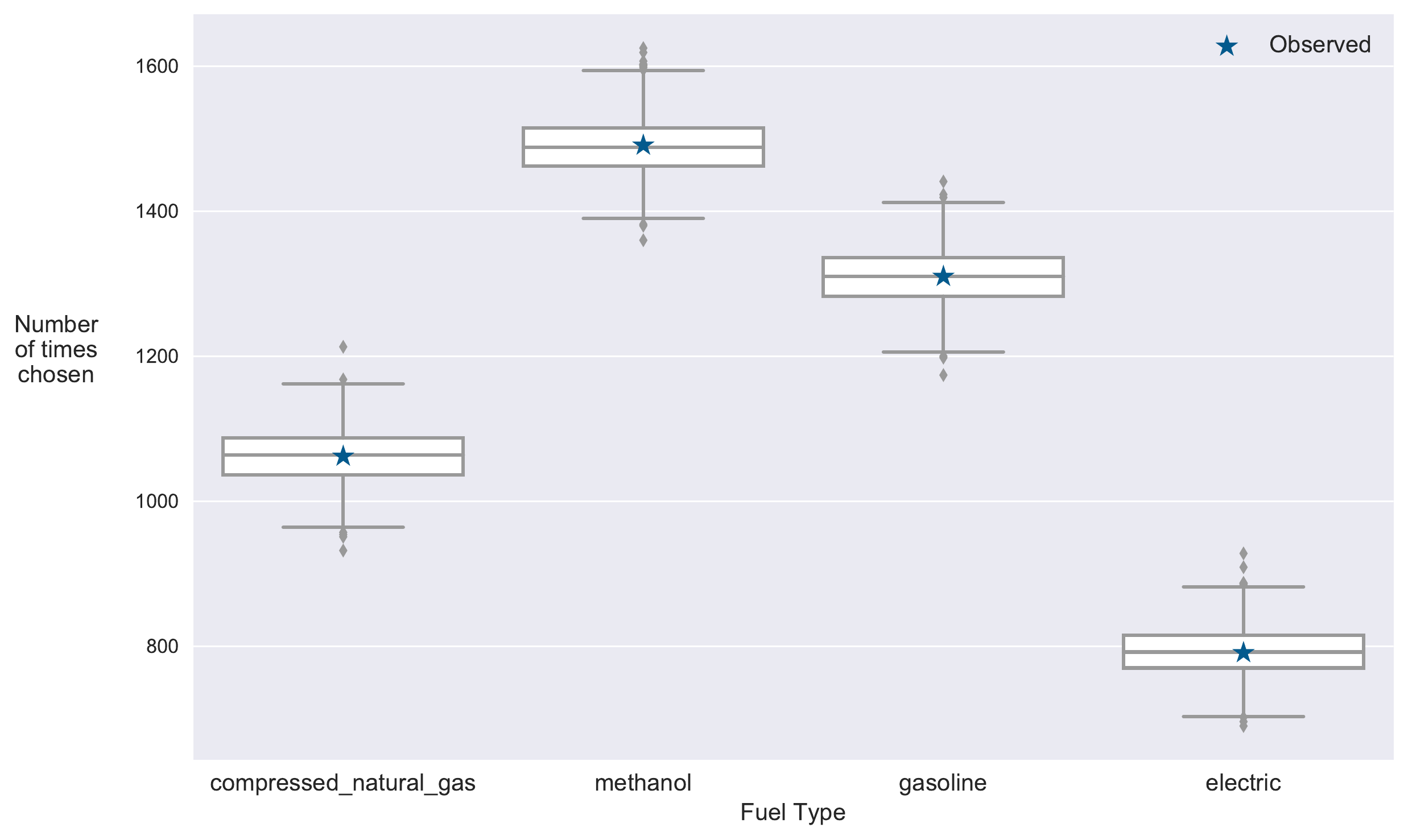}
\caption{Market share plot of \citeauthor{brownstone_forecasting_1998}'s MNL model}
\label{fig:orig-mnl-market-share}
\end{figure}

Market share plots are a more detailed check for lack-of-fit than log-predictive plots. Instead of using a scalar summary, market share plots judge one's model using a vector of numbers, one for each alternative. In particular, market share plots compare the observed versus predicted numbers (or equivalently, percentages) of observations that choose each alternative. See Figure \ref{fig:orig-mnl-market-share} for an example.

Here, the measure of interest is $T \left( Y \right) = \left[ \sum _i y_{i1}, \sum _i y_{i2}, ..., \sum _i y_{iJ} \right]$ where $J$ is the total number of alternatives in one's dataset. As before, $T \left( Y \right)$ is compared against the distribution formed by computing $T \left( Y^r \right)$ for all simulated datasets, indexed by $r$. For this plot, and all others that follow, the simulation procedure is the same as described in Subsection \ref{sec:case1-part1-lp-plots}. 

In general, one can use many different plots to visualize this model check. One can iterate through available alternatives, each time plotting the observed number of decision makers that choose the current alternative versus the predicted distribution of the number of decision makers choosing the current alternative. As with the log-predictive plots, the distribution can be visualized using histograms, kernel density estimates, and/or empirical cumulative distribution functions. The various comparisons may also be concisely displayed using boxplots, as in Figure \ref{fig:orig-mnl-market-share}. Here, the alternatives have been taken to be the various fuel types of each vehicle.

Overall, market share plots assess the degree of underfitting in the marginal distribution, $P \left( Y \right)$, implied by one's model. These plots check whether one's model has learned the aggregate pattern of responses in one's data. Note that such patterns may be of great importance for decision making. As an example, it would be counter-productive to use a vehicle choice model for inventory stocking decisions if the model consistently under-predicted the amount of electric vehicles by a large amount. Even beyond the decision-making use case, it is important to understand the ways that one's model does not fit one's data. For example, if one wants to understand a particular choice process, then market share plots can alert one to the need for caution when interpreting models that frequently under- or over-predict one's alternatives of interest.

As a final caveat, note that a market share plot, as with all the other graphical checks in this paper, is most useful when it ``measures some aspect of the data that might not be accurately fit by the model'' \citep[p.191]{gelman_1996_model}. Practically, this means that if the model being checked ensures that the predicted shares of each alternative match the observed shares, market share plots are not very useful. The most common of such models is an MNL model with $J - 1$ alternative specific constants\footnote{This property is due to the first order conditions of the MNL model with alternative specific constants. See \citet[p.62]{train_discrete_2009} for more details.}. Most other commonly used choice models (e.g. nested logit and mixed logit models) do not ensure that one's predicted market shares match the observed shares \citep{donoso_maximum_2011, klaiber_random_2018}. As a result, market share plots remain a useful tool for checking generalized extreme value (GEV) models, mixed logit models, MNL models without alternative specific constants, and most other models.

\subsubsection{Binned reliability plots}
\label{sec:case1-part1-reliability-plots}
\begin{figure}
\centering
\includegraphics[width=0.75\textwidth]{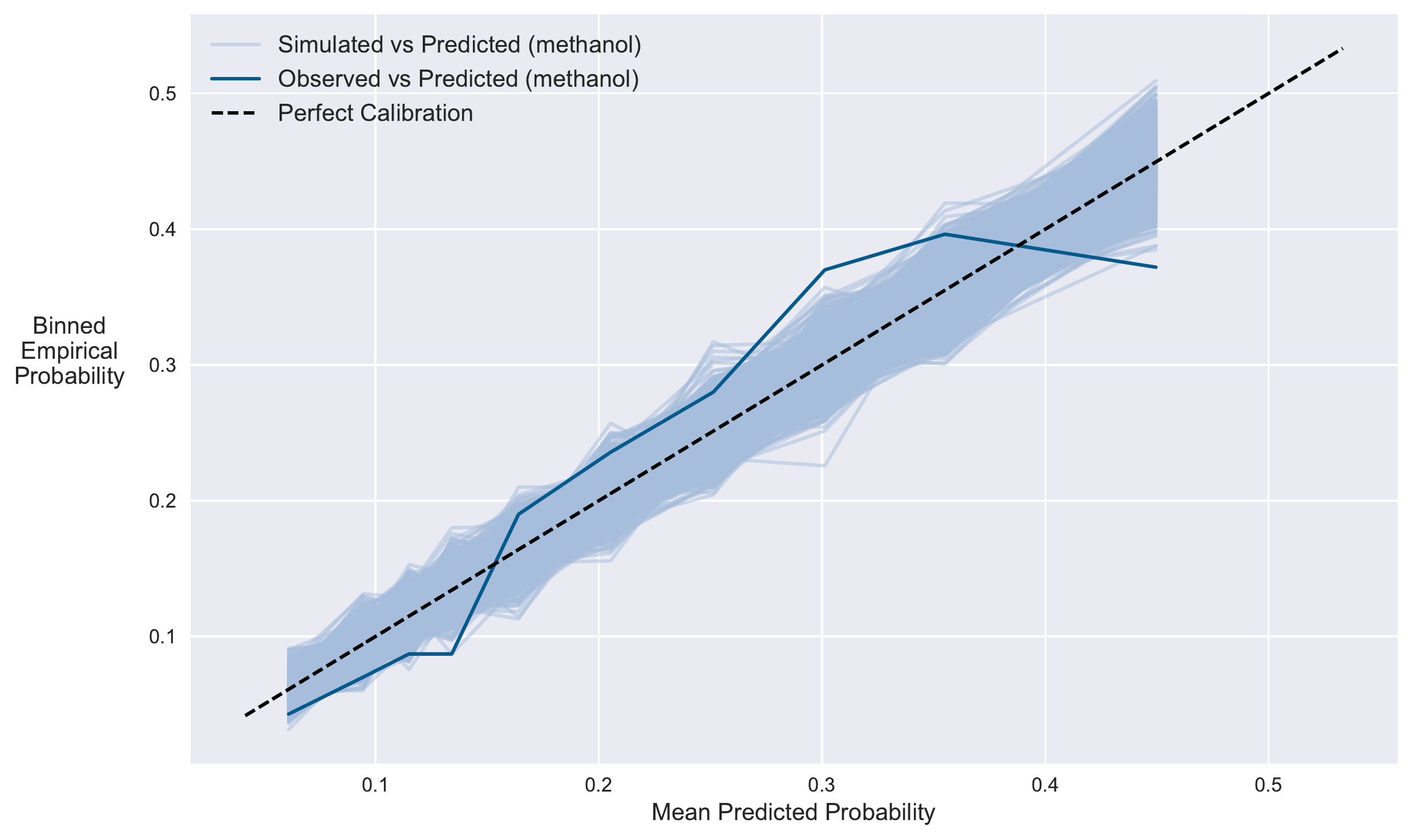}
\caption{Binned reliability plot for methanol vehicles using \citeauthor{brownstone_forecasting_1998}'s MNL model}
\label{fig:orig-mnl-methanol-reliability}
\end{figure}

Conceptually, market share plots judge a model's ability to predict $P \left( Y \right)$ for the various alternatives. In this subsection, binned reliability plots will instead be used to directly evaluate a model's predictions of $P \left( Y | X \right)$, thereby providing a more detailed assessment of a model's predicted probabilities for each alternative. An example of a binned reliability plot for \citeauthor{brownstone_forecasting_1998}'s MNL model is shown in Figure \ref{fig:orig-mnl-methanol-reliability} for methanol vehicles. At its core, binned reliability plots group (i.e. bin) observations according to their predicted probabilities of choosing a given alternative, and then for each group, the mean predicted probability is plotted on the x-axis while the proportion of observations that choose the alternative is plotted on the y-axis. Hopefully, one's observed reliability curve would lie in the middle of the simulated reliability curves and exhibit the same visual trends as the simulated curves. See the following paragraphs for more information and for an explanation of the many principles, choices, and visual elements that comprise this diagram. 

Starting with its name, a reliability plot tries to display the ``reliability'' of a set of probabilistic predictions, where ``reliability refers to the degree of correspondence between forecast probabilities and observed relative frequencies'' \citep[p.41]{murphy_reliability_1977}. If a model is perfectly reliable, then that model's predicted probabilities should equal the ``true probabilities''. As a result, a plot of true probabilities on the y-axis versus predicted probabilities on the x-axis should yield the straight line $y = x$. Of course, one does not observe the true probabilities, one merely observes the discrete outcomes $y_{ij} \in \left\lbrace 0, 1 \right\rbrace \ \forall i, j$. Since discrete choice models generally yield distinct probability predictions for each observation, the sample average of the outcomes at each predicted probability will typically be a poor approximation of the true probabilities. Some smoothing of the observed responses (i.e. averaging outcomes across different predicted probabilities) will therefore be required to yield useful approximations of the desired reliability plot.

On one hand, continuous smooths such as locally weighted scatterplot smoothers \citep{cleveland_bootstrap_1979}, spline smoothers \citep{silverman_spline_1984}, or kernel smoothers \citep{ghosh_kernel_2018} might be preferred since the resulting smoothed curves usually avoid the sharp boundary effects that can occur with ad-hoc grouping. However, I have found such continuous smoothing procedures to be unhelpful for discrete outcome data, especially as the degree of class imbalance increases. As the proportion of observations choosing a given alternative moves away from 50\%, the continuous smooths often led to reliability diagrams that were horizontal lines. This is due to the fact that if a given alternative is unlikely to be chosen, then unless one's model has very high discriminatory power, observations with $y_{ij} = 1$ are likely surrounded by observations with $y_{ij} = 0$. As a result, local averaging (as with continuous smooths) is likely to always yield values near zero, thus producing a horizontal line.

An alternative method of smoothing that avoids the problems just described is based on binning the predicted probabilities. This is the smoothing method used to create the reliability plots in this paper. Using this method, one decides on a number of partitions of the predicted probabilities. Then, the predicted probabilities are ordered from smallest to largest and divided into partitions of roughly equal size. The mean of each partition's predicted probabilities is then plotted on the x-axis and the mean of the outcomes are plotted on the y-axis.

When determining the number of partitions, it should be noted that this choice is subject to a ``precision/robustness'' or ``bias/variance'' trade-off. As the number of partitions increases, one's assessment of the quality of a particular probability prediction is less biased by predicted probabilities of different magnitudes. However, as the number of partitions increases, the variance of the mean outcome within each partition increases due to the reduction in the number of observations being averaged over. The negative effects of having too many partitions is especially acute in class-imbalanced scenarios. Here, each partition's number of observations with $y_{ij} = 1$ may rapidly decrease to zero as the number of partitions increase, thereby destroying one's ability to accurately estimate the ``true'' probabilities. As a result, one can see that increased precision in one's assessments of $P \left( Y \mid X \right)$ comes at the cost of decreased robustness of those assessments. As a general guide, I have empirically found that using ten partitions seems to be adequate for most visualizations\footnote{For example, Figure \ref{fig:orig-mnl-methanol-reliability} uses ten partitions.}. However, I suggest that one experiment with more and less partitions to see whether one's conclusions about the fit of one's model is sensitive to this choice.

Aside from the smoothing method and number of partitions for bin-based smooths, construction of a reliability plot also requires one to choose which predictions will be displayed on the x-axis. For ease of interpretation, I recommend choosing a single set of predicted probabilities for binning and plotting on the x-axis. Such a choice will ensure that each bin is associated with a unique and constant set of observations. With this constant association, one will be able to generate a reference distribution for $T \left( X, Y \right) = \left[ \frac{1}{\mid B_1 \mid} \sum _{i \in B_1} y_{ij} , \frac{1}{\mid B_2 \mid} \sum _{i \in B_2} y_{ij}, \dots, \frac{1}{\mid B_K \mid} \sum _{i \in B_K} y_{ij} \right]$ where $K$ is the total number of partitions, $B_k$ is the set of observations in each partition, and $\mid B_k \mid$ is the number of observations in partition $k \in \left\lbrace 1, 2, \dots, K \right\rbrace$. If multiple sets of predicted probabilities were used (such as one set per simulated  $\beta_r$), then the bin memberships of the observations might change with each draw, thus complicating the description of what we are simulating the reference distribution of.

Now, given that a single set of predicted probabilities is desired, $P \left( Y \mid X, \beta _{\textrm{MLE}} \right)$ is an obvious candidate in a frequentist setting. In a bayesian setting, the point estimate that seems most appropriate is the posterior mean $P_{\textrm{post}} \left( Y \mid X \right) = \int P \left( Y \mid X, \beta \right) P \left( \beta \mid X, Y \right) \partial \beta \approx \frac{1}{R} \sum _{r=1} ^R P \left( Y \mid X, \beta_r \right) P \left( \beta_r \mid X, Y \right)$ where $\beta_r$ is a sample from one's posterior distribution. Either way, once one has chosen a set of predicted probabilities, one can generate a distribution for $T \left( X, Y \right)$ using the simulated $Y^r$ from one's model. One will then be able to compare $T \left( X, Y \right)$ with this reference distribution using a plot such as Figure \ref{fig:orig-mnl-methanol-reliability}. The `observed' curve, i.e. $T \left( X, Y \right)$, will hopefully (1) lie near the line $y = x$, indicating perfect reliability, (2) lie in the middle of the reference distribution, and (3) exhibit the same visual trends as the reference distribution. Deviations from any of these three ideals represent (1) ways that one's model has failed to capture patterns in the data and (2) systematic over- or under-predictions of choice probabilities for the alternative of interest.

\subsubsection{Binned marginal model plots}
\label{sec:case1-part1-mm-plots}
\begin{figure}
\centering
\includegraphics[width=0.75\textwidth]{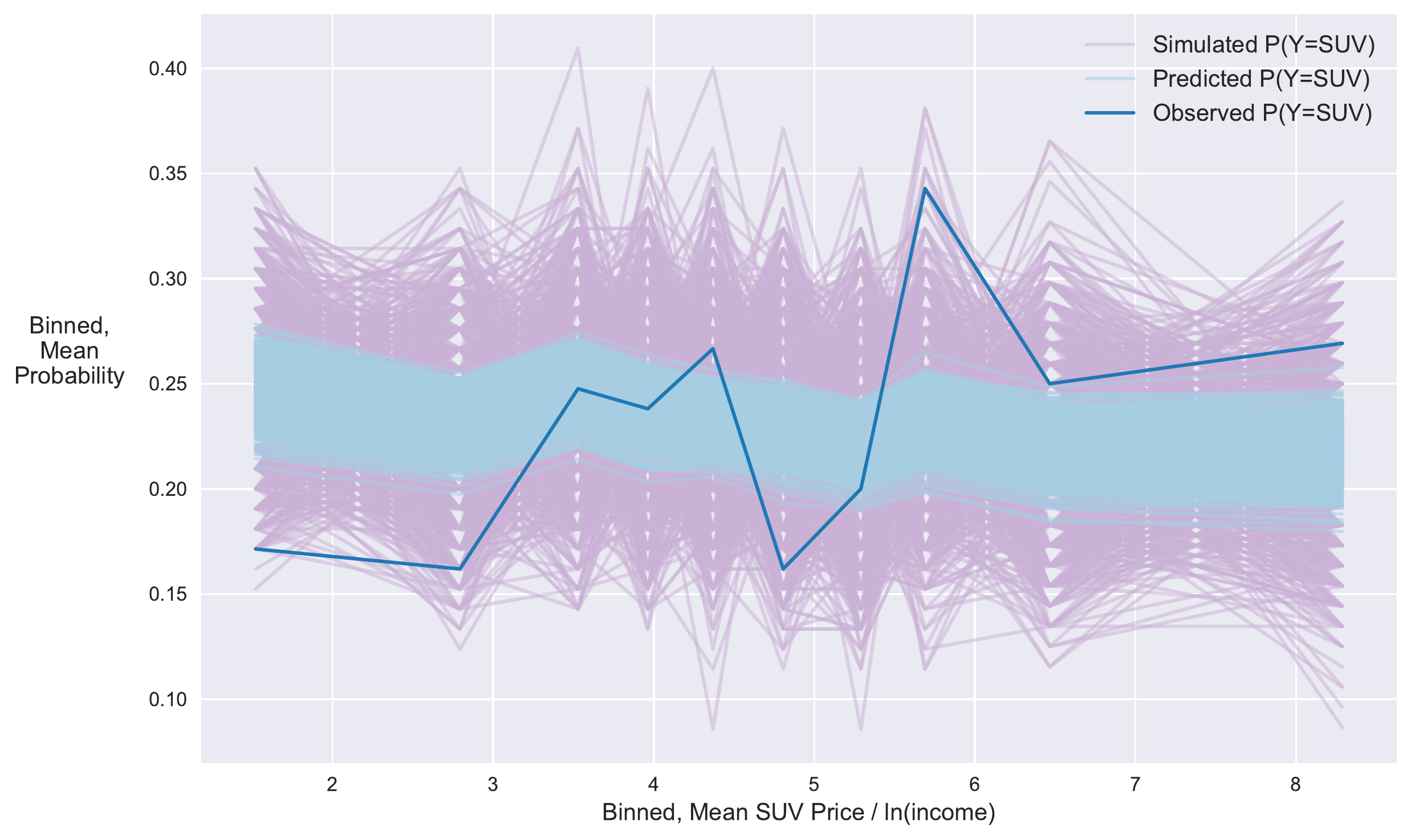}
\caption{Binned marginal model plot for sports utility vehicles using \citeauthor{brownstone_forecasting_1998}'s MNL model}
\label{fig:orig-mnl-suv-marginal}
\end{figure}

Like the binned reliability plots of the last subsection, binned marginal model plots provide a way to check one's model of $P \left( Y \mid X \right)$ for underfitting. Binned marginal model plots display the binned value of a continuous variable on the x-axis and each bin's average probabilities (both observed and simulated) for a given alternative on the y-axis. As a result, this plot checks the degree to which one's model faithfully replicates the empirical relationships between the variable on the x-axis and the true (but unobserved) $P \left( Y \mid X \right)$. For example, in Figure	\ref{fig:orig-mnl-suv-marginal} the binned marginal model plot displays the predicted versus simulated relationship between $P \left( Y \mid X \right)$ and the relative cost of a sports utility vehicle (SUV).

As with reliability plots, the fact that the true probabilities are unobserved means that some smoothing is required. The original marginal model plots \citep{pardoe_2002_graphical} used continuous smoothing techniques such as smoothing splines, but for the reasons discussed in Section \ref{sec:case1-part1-reliability-plots}, bin-based smoothing will be used in this paper. Accordingly, one must therefore determine the number of bins for the plot. To avoid repetition, see the previous subsection for a discussion of considerations for choosing the number of bins. In Figure \ref{fig:orig-mnl-suv-marginal}, 10 partitions were used.

Beyond determining the number of partitions, one must make two additional choices. First, and most importantly, one must decide on the scalar variable to be plotted on the x-axis. Typical candidates to be plotted on the x-axis include the explanatory variables in one's model and variables that are not currently included in one's model. As noted in \citet{pardoe_2002_graphical}, marginal model plots can find variables that add useful information to one's model, even when the p-values of a Wald-test might erroneously suggest such variables' exclusion.

Lastly, one must decide whether to display the averages of the binned, predicted probabilities in addition to the simulated probabilities (i.e. the averages of the binned, simulated choices). For preciseness, note that for each sample ($\beta _r$), one first predicts $P \left( Y \mid X, \beta _r \right)$, and then one uses those probabilities to simulate a vector of choices $Y^r$. The averages of the binned, predicted probabilities are the averages of each bin's $P \left( Y \mid X, \beta_r \right)$, and the averages of the binned, simulated choices are the averages of each bin's $Y^r$. Quantitatively, plotting the averages of the binned, predicted probabilities shows one the relationships that are to be asymptotically expected if one's model is true. Plotting the simulated probabilities shows the sampling distribution (at one's current sample size) of the empirical probabilities, assuming one's model is true.

Personally, I have found that plotting the averages of the binned, predicted probabilities clarifies the relationship that one's model expects asymptotically. For instance, examine Figure \ref{fig:orig-mnl-suv-marginal}. Here, and in general, the averages of the binned, predicted probabilities (shown in light blue) are more tightly clustered than the simulated probabilities (shown in light purple). Quantitatively, this occurs because the averages of the binned, predicted probabilities do not have the inflated variance that the simulated probabilities incur due to the randomness that comes from simulation. Qualitatively, the more tightly clustered curves make it easier to see if one's observed data differs from the (asymptotic) expectations of one's model. These asymptotic judgements can then be qualitatively considered as one uses the binned, simulated probabilities to provide finite-sample assessments of the discrepancies between one's observed data and data generated from one's model. For instance, in Figure \ref{fig:orig-mnl-suv-marginal}, the observed data (shown in dark blue) clearly differs from the asymptotic expectations of \citeauthor{brownstone_forecasting_1998}'s MNL model. This judgement reinforces the fact that the model's simulated choices (of equal size as the observed data) show major differences from the observed data.

\subsubsection{Simulated histograms}
\label{sec:case1-part1-histograms}
So far, the discussed model checking methods have all been aggregate checks of underfitting. To check a model's implied distribution for $P \left( Y \right)$, the log-predictive plots checked underfitting at the level of the entire dataset, and market share plots checked underfitting for each alternative. Becoming more disaggregate, binned, reliability plots and binned, marginal model plots checked for underfitting in a model's predictions of $P \left( Y \mid X \right)$ for specific alternatives using observations that had been grouped according to values of a chosen univariate variable. In the reliability plots, observations were grouped according the their predicted values of $P \left( Y \mid X, \beta _{\textrm{MLE}} \right)$ or $P_{\textrm{post}} \left( Y \mid X \right)$. For the binned, marginal model plots, observations were grouped according to whatever variable was chosen for the x-axis.

\begin{figure}
\centering
\includegraphics[width=0.75\textwidth]{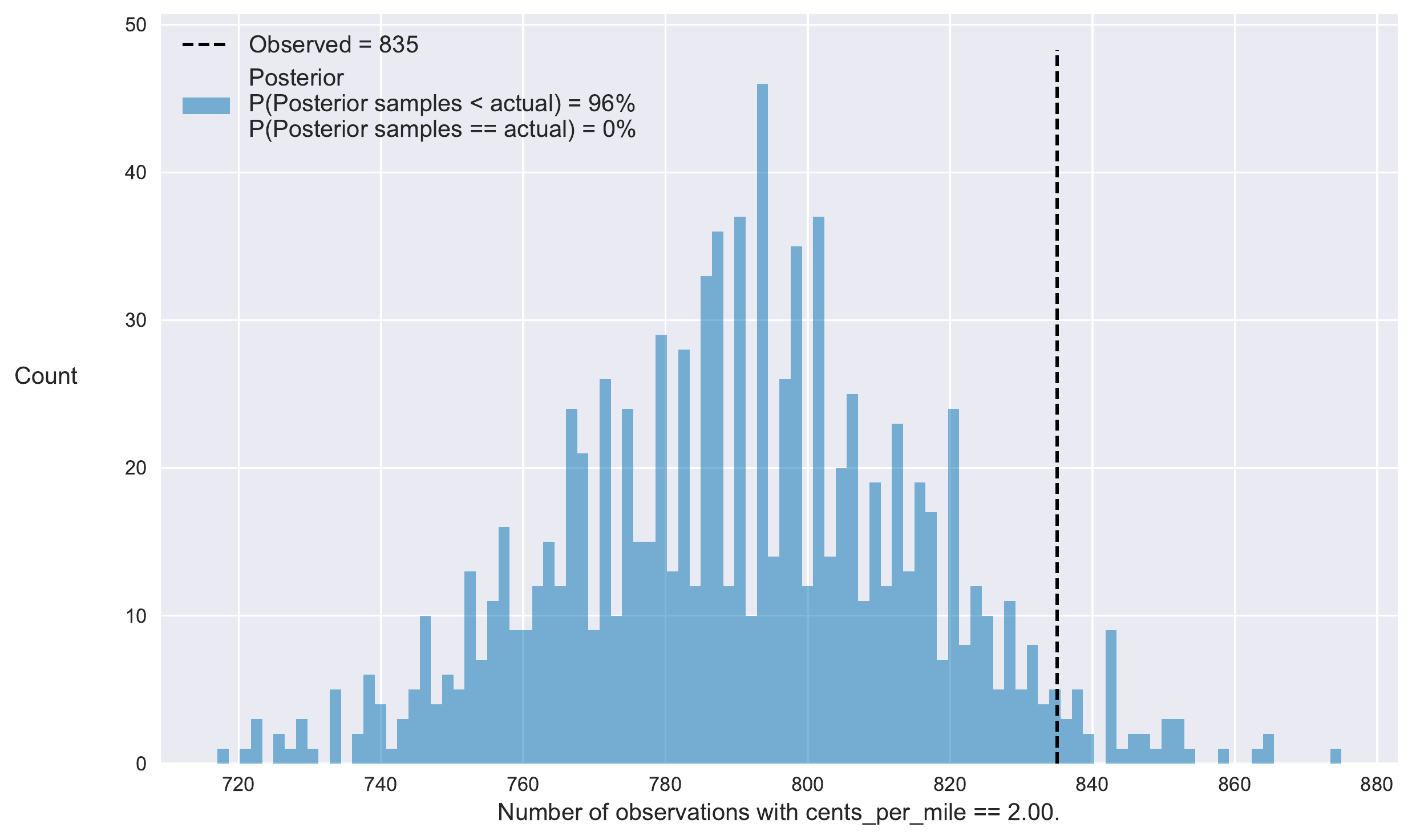}
\caption{Simulated histograms for regular cars using \citeauthor{brownstone_forecasting_1998}'s MNL model}
\label{fig:orig-mnl-regcar-histogram}
\end{figure}

In this subsection (and the following two subsections), I introduce this paper's most granular model checking methods. These techniques judge a model's implied distribution of $P \left( X \mid Y \right)$ for underfitting of specific alternatives at each observed value of a chosen univariate variable. Here, I introduce simulated histograms to assess underfitting of a discrete explanatory variable. Figure \ref{fig:orig-mnl-regcar-histogram} provides a concrete example. In this plot, I assess \citeauthor{brownstone_forecasting_1998}'s MNL model with respect to its ability to predict the number of choice observations that choose regular cars with an operating cost of 2 cents per mile. From the plot, we see that while 835 observations actually chose a regular car with an operating cost of 2 cents per mile, 96\% of the simulated datasets had less than 835 observations choosing regular cars that operate at 2 cents per mile. The observed dataset is therefore unusual with respect to this feature, and the simulated histogram displays the model's lack-of-fit at an operating cost of 2 cents per mile for regular cars.

Generalizing beyond this specific example, one constructs a simulated histogram by first choosing a given criteria. The number of observations meeting this criteria will be one's `discrepancy' measure, $T \left( X, Y \right)$. Then, this discrepancy measure is tabulated in both the observed dataset and in each of one's simulated datasets. Next, the number of simulated observations meeting the criteria is plotted in a histogram to show the distribution of the discrepancy measure, assuming that one's model is correct. Finally, one draws a vertical line on the plot to indicate the value of the observed discrepancy $T \left( X, Y \right)$. For the simulated histograms in this paper, the discrepancy measure is the number of observations that choose a given alternative when the discrete variable of interest takes on some specified value. As an example, in Figure \ref{fig:orig-mnl-regcar-histogram}, the alternative is regular cars; the variable of interest is the operating cost of the vehicle, and the specified value is 2 cents per mile.

\subsubsection{Simulated kernel density estimates}
\label{sec:case1-part1-kdes}
\begin{figure}
\centering
\includegraphics[width=0.75\textwidth]{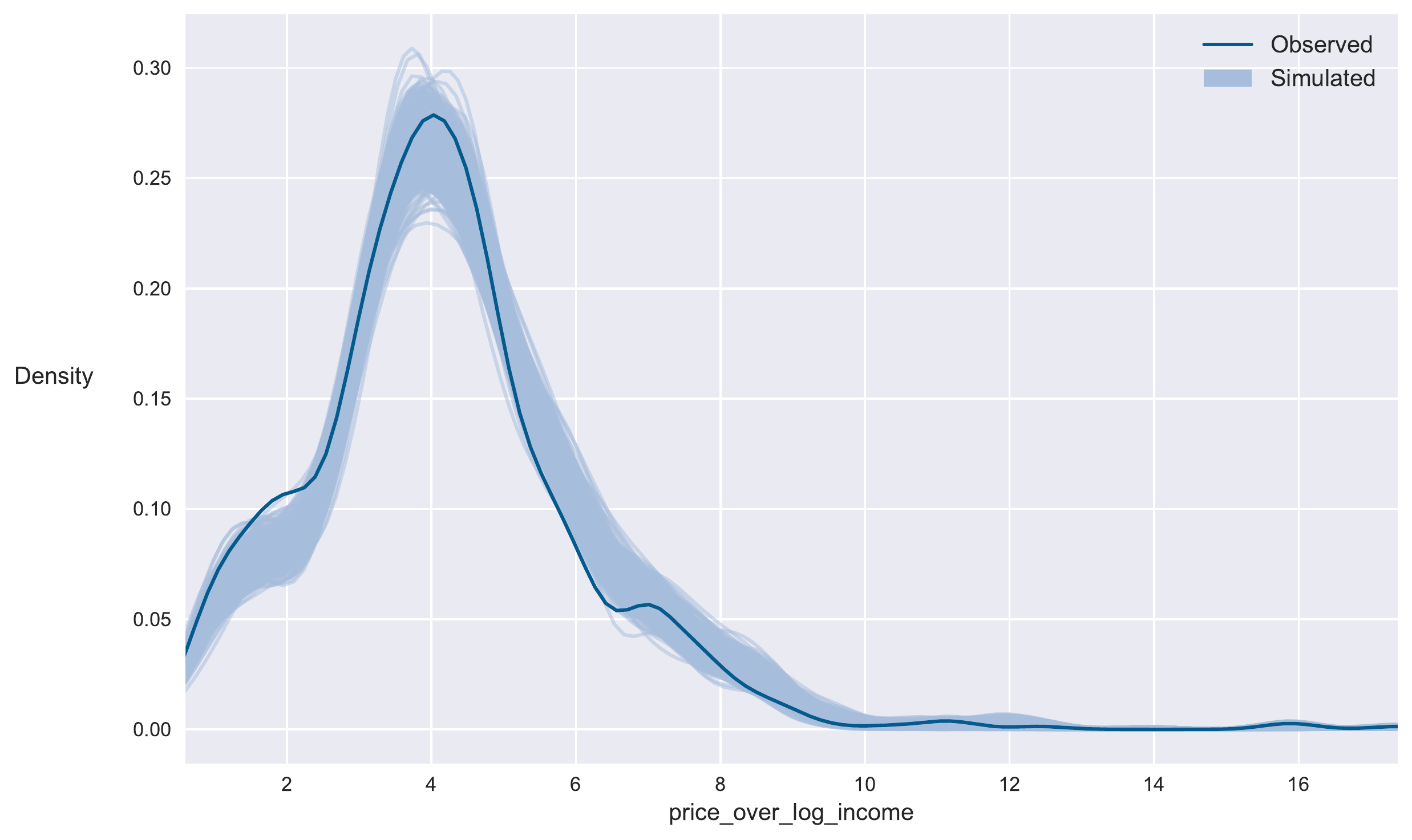}
\caption{Simulated kernel density estimates for electric cars using \citeauthor{brownstone_forecasting_1998}'s MNL model}
\label{fig:orig-mnl-electric-kde}
\end{figure}

In the last subsection, simulated histograms were used to judge underfitting for specific alternatives at each observed value of a discrete variable. This section introduces simulated kernel density estimates\footnote{Note that kernel density estimates are non-parametric estimates of the probability density function of a random variable. For an econometrician's and statistician's introduction to KDEs, see \citet{racine_nonparametric_2008} and \citet{wasserman_all_2006}, respectively.} (KDEs): an analogous model checking method that assesses underfitting for specific alternatives with respect to the individual values of a continuous variable. Figure \ref{fig:orig-mnl-electric-kde} presents an example of simulated KDEs. 

In this example, the alternative of interest is electric vehicles and the variable of interest is the relative price: the price of the electric vehicle, divided by the natural log of the purchaser's household income. As a result, Figure \ref{fig:orig-mnl-electric-kde} plots the kernel density estimates of the distribution of relative prices for decision makers that were observed or simulated to have chosen an electric vehicle. From the plot, we see that (1) around a relative price of 2, the observed KDE strays outside the band of simulated KDEs; (2) between a relative price of approximately 5 and 6.5, the observed KDE drifts to the bottom edge of the simulated KDEs; and (3) between a relative price of 6.5 and 7, the observed KDE is mostly flat whereas the simulated KDEs maintain a negative slope. As noted in Section \ref{sec:case1-part1-reliability-plots}, one wants the curves generated by the simulated data to envelop and straddle the curve generated by the observed data, and one wants the curves generated by the simulated data to display the same visual trends as the curve generated by the observed data. All three points mentioned above highlight values of the relative price where the simulated KDEs fail to envelop, straddle, or mimic the observed KDE. I.e., the simulated KDE plot reveals multiple ways in which \citeauthor{brownstone_forecasting_1998}'s MNL model fails to capture patterns in the observed relationship between relative price and the choice of electric vehicles.

Now, to construct a simulated KDE plot as used in this paper, the procedure is as follows. First, choose a continuous, univariate variable to be plotted on the x-axis. Next, choose the alternative that the KDEs will be plotted with respect to. Given the selected alternative and variable, construct the plot by iterating through each of the simulated vectors of choices ($ Y^r$) and the observed choices ($Y$). For each set of choices, select the observations that choose the alternative of interest, and from those selected observations, plot a KDE of the chosen variable. Note that in plotting the KDE of the chosen variable, one must choose a kernel and bandwidth for the KDE. In practice (and in Figure \ref{fig:orig-mnl-electric-kde}), I have found that the standard normal distribution kernel and Scott's (\citeyear{scott_multivariate_1992}) default bandwidth procedures are typically adequate for visualization purposes. Researchers are of course free, and encouraged, to experiment with alternative kernels and bandwidth selection procedures.

\subsubsection{Simulated cumulative distribution functions}
\label{sec:case1-part1-cdfs}
\begin{figure}
\centering
\includegraphics[width=0.75\textwidth]{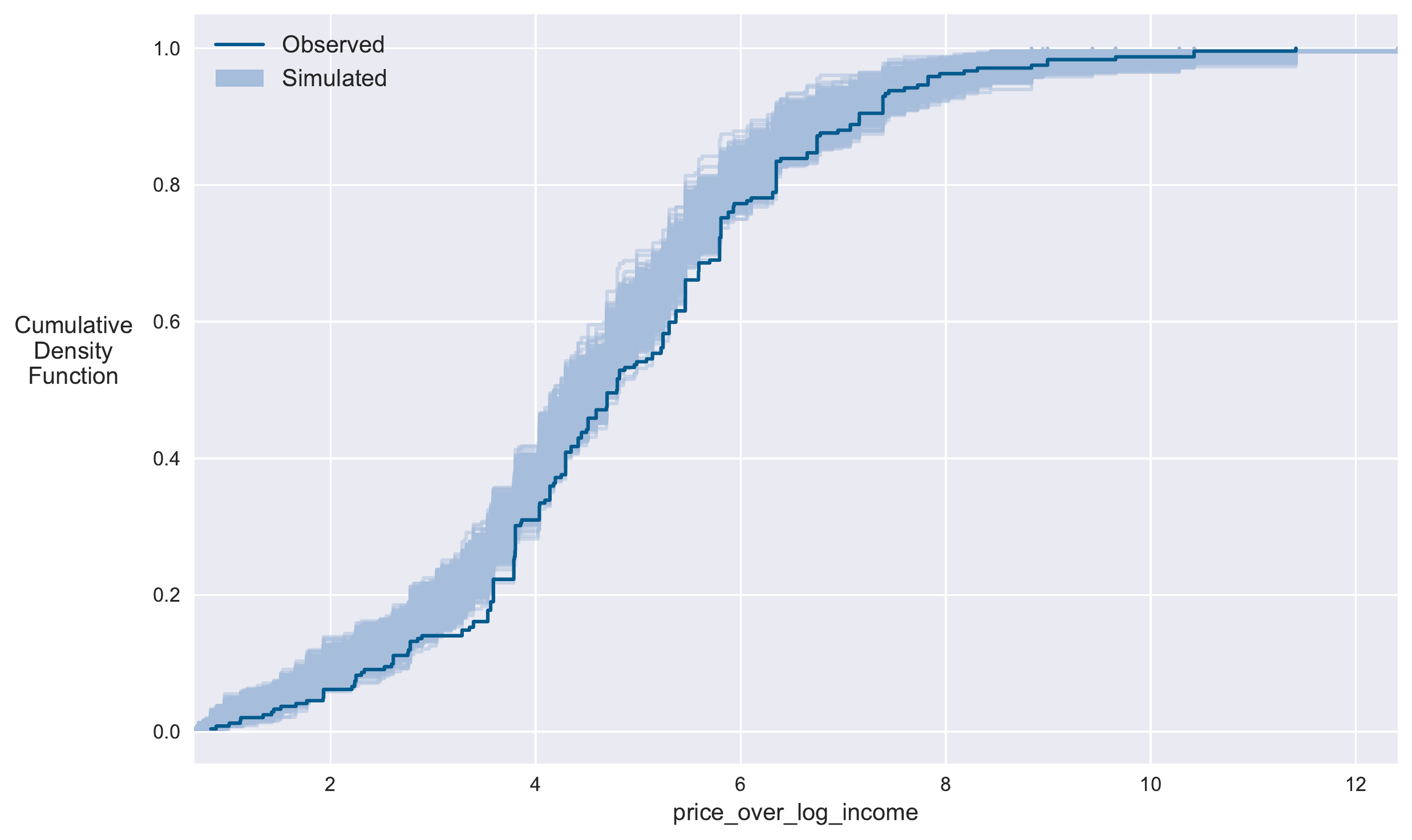}
\caption{Simulated CDFs for sport utility vehicles using \citeauthor{brownstone_forecasting_1998}'s MNL model}
\label{fig:orig-mnl-suv-cdf}
\end{figure}

This paper's final type of plot is a plot of simulated cumulative distribution functions (CDFs). Conceptually, simulated CDFs serve the same role as simulated KDEs: by visualizing $P \left( X \mid Y \right)$, these plots are used to assess underfitting for specific alternatives with respect to the individual values of a continuous variable. Figure \ref{fig:orig-mnl-suv-cdf} gives an example of a simulated CDF plot.

As in Section \ref{sec:case1-part1-mm-plots}, the variable of interest is again the relative price, and the alternative of interest is again SUVs. This time, however, observations will not be binned according to their relative price values. Instead, for each vector of choices (both simulated and observed), the observations that choose SUVs will be sorted in ascending order of their relative price values. These relative prices will be plotted on the x-axis. Then, for each relative price (for the current vector of choices), the y-axis value will be the proportion of observations that chose SUVs and have relative prices less than or equal to the x-axis value. As always, we would like the CDF generated by the observed data to lie in the center of the CDFs generated by the simulated data, and we would like the simulated CDFs to display the same visual trends as the observed CDF. In the example given by Figure \ref{fig:orig-mnl-suv-cdf}, the observed CDF is clearly at the bottom edge or outside the band of simulated CDFs for most of the observed relative price values. This indicates that \citeauthor{brownstone_forecasting_1998}'s MNL model is underfit with respect to the relative price variable and SUVs.

Now, given that simulated CDFs and simulated KDEs  perform the same conceptual role in assessing underfitting, why should both types of plots be used? The answer is that the two types of plots are complementary. For instance, in Section \ref{sec:case1-part1-kdes}, I noted that one has to choose a bandwidth and kernel in order to construct the simulated KDEs. Simulated CDF plots require no such ad-hoc choices. Moreover, while simulated KDEs highlight the exact values where one's model expect more or less observations (i.e. where one's model predicts a higher or lower density) than observed in one's data, it can be difficult to visually identify situations when one's model has consistently but only slightly over- or under-predicted the density for a sequence of values. Such situations are easily identified with a simulated CDF plot. Conversely, while simulated CDFs excel at exposing underfitting, simulated KDEs are potentially more useful for fixing it. Indeed, knowing a variable's density with respect to a given alternative can suggest helpful transformations of that variable in one's systematic utility functions \citep{kay_transformations_1987, bergtold_bernoulli_2009}.

Lastly, the particular set of simulated CDFs in Figure \ref{fig:orig-mnl-suv-cdf} may seem redundant given that the relationship between the choice of SUVs and relative price was assessed in Figure \ref{fig:orig-mnl-suv-marginal}. Again, these two plots are found to be complementary. The simulated CDF plot does not bin any observations, so we do not lose any information, and we are able to see that the relationship between relative price and SUVs is poorly captured for all relative prices below about 7.5. Because the simulated curves in Figure \ref{fig:orig-mnl-suv-marginal} do not exhibit the same visual trend as the observed curve, the marginal model plot also indicates the lack-of-fit in the MNL model's estimated relationship between the relative price and the choice of SUV. However, Figure \ref{fig:orig-mnl-suv-marginal} also points out that the partitions with predicted probabilities between approximately 0.20 and 0.27 are actually predicted accurately. This highlights the fact that at least some of the predicted probabilities of choosing SUV are trustworthy, even if the overall estimated relationship is not. Depending on one's use case, this may or may not be sufficient.

\subsubsection{Discussion}
\label{sec:case1-part1-discussion}
In Subsections \ref{sec:case1-part1-lp-plots} - \ref{sec:case1-part1-cdfs}, I presented seven graphical checks, five of which (Figures \ref{fig:orig-mnl-methanol-reliability} - \ref{fig:orig-mnl-suv-cdf}) revealed patterns of underfitting in \citeauthor{brownstone_forecasting_1998}'s MNL model. In particular, the reliability plot in Figure \ref{fig:orig-mnl-methanol-reliability} reveals that the MNL model's probability predictions for methanol vehicles are generally unreliable. Analogous plots\footnote{These plots are not shown due to space considerations. However, they are available upon request and at \url{https://github.com/timothyb0912/check-yourself/blob/master/notebooks/_06-tb-Make_introductory_plots.ipynb}.} for the other vehicle fuel and body types show that this pattern of inaccurate probability predictions is consistent across multiple dimensions. The conclusion from these plots is that overall, \citeauthor{brownstone_forecasting_1998}'s MNL model underfits numerous aspects of the observed data.

By itself, the fact that the MNL model does not adequately represent the data generating process is unsurprising. Indeed, \citet{brownstone_forecasting_1998} began their paper with the presumption that the MNL model would be overly simple and would be better replaced with models such as the mixed logit and probit models. Instead, the checks in Figures \ref{fig:orig-mnl-suv-marginal} - \ref{fig:orig-mnl-suv-cdf} are most useful for the model-free way that they reveal lack-of-fit in the MNL model. Specifically, without relying on comparisons between two potentially flawed models, Figures \ref{fig:orig-mnl-suv-marginal} - \ref{fig:orig-mnl-suv-cdf} show that \citeauthor{brownstone_forecasting_1998}'s MNL model does not adequately relate individual choices to relative costs or operating costs for the various vehicle fuel and body types. All together, these checks allow one to directly assess whether key relationships have been captured in one's model. As a result, model assessment is effectively separated from the (arguably less-important) task of model comparison: a task that often reduces to checking whether one potentially flawed model has a predictive score (e.g. log-likelihood) that is `sufficiently' higher than another potentially flawed model.

\subsection{General Methodology}
\label{sec:general-methodology}
Now that Section \ref{sec:case1-part1} has introduced specific examples of this paper's model checking techniques, the general methodology can be introduced clearly.

In particular, the model checking methodology based on predictive simulations is as follows. First, let the observed data be represented by $\left( X, Y \right)$. Note that not all of the variables in $X$ need be part of one's choice model specification. Next, specify one or more statistics of the data. For each statistic, $T$:
\begin{enumerate}
\item Calculate $T \left( X, Y \right)$ for the observed data, yielding a value $T_0$. Note that $T_0$ may be a vector, e.g. an empirical cumulative density function.

\item From some distribution, draw $R$ samples of one's model parameters\footnote{Note, we sample from a distribution of parameters because we are interested in assessing our model, not in assessing a particular point estimate for our model. We draw parameter samples to account for our uncertainty about the true parameter value if one is a bayesian or about the possible datasets if one is a frequentist.}. For bayesian analyses, the distribution could be one's prior or posterior distribution (depending on whether one wants to check the prior or posterior). For frequentist analyses, the distribution may be the sampling distribution of one's model. The sampling distribution can be approximated using the asymptotic sampling distribution or by using the paired bootstrap \citep{brownstone_bootstrap_2001}.

	\begin{enumerate}
		\item For each sampled vector of parameters $\left( \beta _r \right)$, calculate the predicted probabilities $\left( P \left[ Y \mid X, \beta_r \right] \right)$ of each available alternative being chosen by the decision makers in one's dataset.

		\item For each set of predicted probabilities, simulate a choice for each decision maker.

		\item For each set of simulated choices $\left( Y^r \right)$, calculate $T \left( X, Y^r \right)$.
	\end{enumerate}

\item Compare $T_0$ with the distribution formed by the samples $T \left( X, Y^r \right) \forall \  r \in \left\lbrace 1, 2, ..., R \right\rbrace$.

\item The more  `surprising' $T_0$ is given one's simulated values of $T$, the more one's model underfits one's data (in terms of the statistic $T$).
\end{enumerate}

As examples, I specified $T$ as a function with vector outputs for the simulated KDEs and CDFs; for the binned, marginal model plots and binned, reliability plots; and for the market share plots. Conversely, I specified $T$ as having scalar outputs for the simulated histograms and log-predictive plots.

Regardless of how one specifies $T$, the model-checking procedure described above requires an analyst to choose one or more statistics $T$ to be checked. In general, researchers should always specify statistics that are of substantive importance in one's study. For example, if building a bicycle mode choice model using revealed preference data, one might check the observed number of cyclists in different geographic regions of one's study, even if one had not explicitly built a `spatial choice model.'

However, despite the admonition to thoughtfully define the statistics that are being checked, I readily acknowledge that analysts may not always have a set of statistics in mind. Moreover, it is useful to know, in many specific but common ways, whether or not one's model fits one data. To this end, I propose the following new method for ``automatically'' checking the fit of one's model to one's data.

\begin{enumerate}
\item In the case of ``labeled'' alternatives (e.g. shopping brand or travel mode choice) choose one or more alternatives of interest.
\label{step:labeled-case}

	\begin{enumerate}
	\item For each alternative of interest, construct market share plots and binned reliability plots to perform a basic assessment one's probability predictions.
	\item For each alternative of interest, cycle through the related, available, explanatory variables. For instance, continuing with bicycle mode choice, if one is looking at cyclists, then examine bicycle travel time instead of automobile travel time.
	\item If the explanatory variable being examined is discrete, iterate through the distinct values of the variable.
	\item For each value of the discrete explanatory variable being examined, define $T$ to be the number of individuals with the current value of the current explanatory variable that choose the current alternative.
	\item If the explanatory variable being examined is continuous, then define $T$ to be the empirical cumulative density function or a kernel density estimate of the explanatory variable, conditional on choosing the current alternative.
	\end{enumerate}
	
\item In the case of ``unlabeled'' alternatives,(e.g. travel route choice or vehicle choice), if one has a discrete explanatory variable of great importance (e.g. vehicle type in a vehicle choice model), then treat the values of that discrete explanatory variable as one's ``alternative labels'' and proceed as in Step \ref{step:labeled-case}.

\item In the case of unlabeled alternatives without a discrete explanatory variable of great importance, one can treat the ``labels'' as ``chosen or not'', and define the alternatives of interest to be the ``chosen'' alternatives. Then proceed as in Step \ref{step:labeled-case}.
\end{enumerate}

This proposed method is \textit{nearly} automatic because it merely requires one to define ``proxy labels'' for unlabelled alternatives and to define the variables of interest that are related to each distinct alternative. Note that since random utility models are the most common type of discrete choice model, each alternative's systematic utility function is likely to specify which variables relate to which alternatives.

\section{Relations to Existing Literature}
\label{sec:lit-review}
Section \ref{sec:methods} introduced seven specific model checks and a general method for using predictive simulations to assess underfitting in one's discrete choice model. In this section, I clarify how my proposed methods deliver increased benefits compared to pre-existing model checking and model comparison techniques. First, Section \ref{sec:review-graphical-checks} shows how my proposed methods continue the progression of graphical model checks in statistics from residual plots to posterior predictive checks. Here, I highlight how my proposed methods facilitate the routine use of checks for underfitting and how they offer new ways to check one's model specifications. Next, Section \ref{sec:review-model-comparison} contrasts my proposed methods with model comparison techniques---the more common class of methods in discrete choice. In this subsection, I explain how my proposed model checking techniques avoid the limitations imposed by model comparison methods when checking for underfitting.

\subsection{Graphical Model Checks}
\label{sec:review-graphical-checks}
Historically, much research has focused on the development of graphical methods for checking predictive models. Dating back to at least the 1960s \citep{anscombe_examination_1963, cox_general_1968}, the most widely known graphical model checks are residual plots for linear regression. Conceptually, residual plots are similar to the plots described in Section \ref{sec:methods}: they implicitly compare an observed vector of residuals against the vectors that would be expected if one's model was true. Specifically, if one's data are truly generated by one's model, then under common linear regression assumptions\footnote{For e.g., the assumptions of the Gauss-Markov theorem.}, one's residuals (i.e. the observed data minus one's predictions) are independently and identically distributed according to a normal distribution with mean zero and constant variance. Accordingly, one's residual plots would be expected to look like a series of random draws from a (scaled) standard normal distribution. This simplicity of the residual plot's expected patterns enabled one, without simulation, to visually determine when the observed residuals were markedly different from residuals that could be generated by one's model. Depending on the type of residual plots being examined, marked differences between expected and observed patterns of residuals could then be used to identify when and how one's linear function was misspecified, to identify when transformations of the outcome variable would be useful, to identify when major assumptions such as independence across observations were violated, etc. \citep{anscombe_examination_1963, cox_general_1968}.

Following the tremendous success of graphical model checks for linear regression, and encouraged by the unifying framework of generalized linear models, researchers in the 1980's sought to extend graphical methods like residual plots to discrete choice models (e.g. binary and multinomial logistic regression). Unfortunately, such efforts were frustrated by the discrete dependent variables inherent to discrete choice models \citep[pp. 747-748]{albert_bayesian_1995}. In particular, discrete dependent variables complicate the expected patterns in one's residual plots, thereby precluding easy, implicit assessment of one's model. To cope with these challenges, even the earliest efforts leveraged many modern features of graphical model checks. For example, \citet{landwehr_graphical_1984} is one of the first papers to introduce graphical checks for underfitting in (binary) discrete choice models. As with my proposed methods in Section \ref{sec:methods}, \citet[p. 64]{landwehr_graphical_1984} use predictive simulations to create reference distributions that are explicitly displayed on one's plots. Moreover, \citeauthor{landwehr_graphical_1984} (p. 66) also use smoothing to deal with the discrete nature of the quantities being plotted. Even the idea of using the observed distributions of $P \left( X \mid Y \right)$ to check one's model was already noted in Donald B. Rubin's discussion (pp. 79 - 80) of \citet{landwehr_graphical_1984}. Though differing from Rubin in implementation, his idea directly inspired my creation of the simulated histograms, KDEs, and CDFs.

In summary, the ``classical'' diagnostic plots for discrete choice models compare and contrast with this paper's proposed techniques as follows. As a point of commonality, the classical plots pioneered many of the techniques upon which my proposed methods rest: using predictive simulations to construct reference distributions, using smoothing to plot discrete quantities, and checking $P \left( X \mid Y \right)$ as well as $P \left( Y \mid X \right)$. However, research on classical diagnostic plots was more limited than this paper's proposed techniques. In particular, classical diagnostic plots were produced in a piecemeal fashion: one stroke of scholarly genius at a time. No unifying connections were made between these few distinct sets of graphical checks, and no unifying methodology was proposed for creating tailored graphical checks for one's unique modeling problems.

For example, \citeauthor{murphy_reliability_1977} (\citeyear{murphy_reliability_1977}, \citeyear{murphy_general_1987}) introduced reliability plots to judge one's probability predictions, and a reference distribution based on predictive simulations was developed for these plots in \citet{brocker_increasing_2007}. In the intervening years between these efforts, \citet{landwehr_graphical_1984} introduced empirical probability plots, local mean deviance plots (a graphical form of subgroup analyses), and partial residual plots. Here, empirical probability plots (indirectly) assess the same relationship as reliability plots, but no connection between the two techniques was made\footnote{For completeness, note that reliability plots can be considered graphical representations of the numeric Hosmer-Lemeshow goodness-of-fit statistic \citep{lemeshow_review_1982}. Reliability plots are also similar in spirit to the Horowitz-Louviere test \citep{horowitz_testing_1985, horowitz_testing_1993} in the sense that they directly compare observed choices versus predicted probabilities.}. Finally, model checking graphs such as the ``Separation Plot'' \citep{greenhill_separation_2011} and the ``Heat Map Plot'' \citep{esarey_assessing_2012} have been recently created by researchers in quantitative political science. Of particular interest is the heat map plot which is based on familiar techniques such as smoothing and creating reference distributions based on predictive simulations. Unfortunately, although this plot is the exact analogue of the reliability plot (simply using continuous smoothing instead of binning), the connection between the two types of plots has been ignored.

In contrast to these classical diagnostic plots, a second generation of graphical model checks has emerged since the mid-1990s. Beginning in earnest with the \textit{posterior predictive checks} of \citet{gelman_posterior_1996}, this framework unifies the various model checking plots. For example, building on the approaches described in \citet{gelman_exploratory_2004} and \citet{buja_statistical_2009}, all the graphs in Section \ref{sec:case1-part1} correspond to different choices of a test statistic $T$ and different ways of plotting the reference distribution that is created by the predictive simulations. Aside from these two choices, graphical model checks basically follow the seven-step procedure detailed in Section \ref{sec:general-methodology}. This clarity and abstractness of the procedure for creating graphical model checks has numerous benefits. For instance, the abstractness of the procedure has allowed bayesians and frequentists to make use of a single methodology, merely substituting their own preferred methods for sampling from their estimated models. Moreover, by appropriately defining $T$, the common methodology makes it easy to define custom checks that are tailored to one's modeling context.

Clearly, the model checking techniques proposed in this paper draw heavily upon the posterior predictive checking framework. For instance, \citet[p. 896]{evans_checking_2006} note that
\begin{quote}
``[...] the prior induces a probability distribution on the set of possible likelihood functions via the prior predictive probability measure $M$. If the observed likelihood [...] is a surprising value from this distribution, then this would seem to indicate that a prior-data conflict exists.''
\end{quote}
The log-predictive plots merely re-purpose this observation by substituting the posterior distribution for the prior distribution. Similarly, the market share plots are simply graphical depictions of the market segment prediction tests of \citet[Section 7.6]{ben-akiva_discrete_1985}, alongside a reference distribution created by predictive simulations. Continuing through the remaining plots, the binned reliability plots are the same as their classical counterparts, but enhanced with a graphical depiction of the reference distribution created by the predictive simulations from one's model. Marginal model plots are explicit descendants of the posterior predictive checking framework \citep[pp.266-267]{pardoe_2002_graphical}, and this paper has only modified these plots to use bin-based smoothing instead of continuous smooths. Lastly, I noted above that as far back as 1984, Donald Rubin [in \citet[pp. 79-80]{landwehr_graphical_1984}] suggested that
\begin{quote}
``[...] for diagnostic checking of estimates of $\textrm{Pr} \left( y = 1 \mid x \right)$, my preference is reversed. I believe that we should examine the consequences of a proposed estimate of $Pr \left( y = 1 \mid x \right)$ using the discriminant analysis formulation [i.e. $\textrm{Pr} \left( y = 1 \mid x \right) \propto \textrm{Pr} \left( x \mid y = 1  \right) \textrm{Pr} \left( y = 1 \right)$].''
\end{quote}
My proposed plots of simulated histograms, KDEs, and CDFs are merely new applications of Rubin's insight to examine the implied class-conditional distributions $P \left( X \mid Y \right)$, complemented by reference distributions that are simulated from one's estimated model.

Given that the plots of Section \ref{sec:case1-part1} are firmly grounded in the posterior predictive framework, some readers may wonder if there is anything truly new in this paper's proposed methodology. There is. Thus far, practitioners who use posterior predictive checks have restricted themselves to developing custom graphical checks for each new choice modeling endeavor. See \citet{gelman_diagnostic_2000} and \citet{carlin_case_2001} for examples from statisticians, and see \citet{jedidi_measuring_2003}, \citet{musalem_coupon_2008}, and \citet{gilbride_posterior_2010} for examples from the discrete choice literature. Except for \citet{jedidi_measuring_2003}, the checks in these papers are mostly specific to their particular studies/models and would not be immediately applicable to analysts in general.

In Section \ref{sec:general-methodology}, however, I proposed a new approach for automatic model checking of discrete choice models. Compared to the existing literature, my model checking recommendations are more broadly applicable and more comprehensive. First, my proposed methods are applicable whenever one has disaggregate choice data available. The applicability of my methods does not depend on the specific choice model being used. Secondly, my proposed checks are more comprehensive than checks provided in most other papers because they assess model fit on multiple dimensions. They allow one to assess the fit of one's model overall (using log-predictive plots), on particular alternatives (using market share plots and binned reliability plots), and on particular variables (using binned marginal model plots, simulated histograms, simulated KDEs, and simulated CDFs).

Of course, I am not claiming that my proposed model checks account for all types of checks that one may wish to do. For references to other model checks that may be of use, see the last paragraph of Appendix A. Moreover, I do not make the claim that my proposed methods will always uncover an underfit model, nor that my proposed methods are the best automatic model checking procedures that one can follow. Finally, I do not think there is any problem with using model checking techniques that are tailored to one's problem and chosen modeling technique. On the contrary, I believe that such targeted model checks are complementary to the checks described in this paper. I simply note the lack of commonly-used or agreed-upon checks for discrete choice models, and I propose my methodology as one useful approach to facilitate the routine use of graphical model checks for underfitting.

\subsection{Model Comparison versus Model Checking}
\label{sec:review-model-comparison}
Thus far, most of the articles cited in this literature review have been examples of graphical model checking from the statistics literature. This is by necessity as opposed to choice. Overall, there have been few examples of graphical model checking within the discrete choice literature. The papers of \citet{dunn_graphical_1987} and \citet{nagel_diagnostics_1992} are notable exceptions. Instead, most model diagnostics within the discrete choice literature are instances of model comparison as opposed to model checking.

For concreteness, take ``A Diagnostic Test for the Multinomial Logit Model'' by \citet{tse_diagnostic_1987}. This diagnostic test does not directly compare the observed data with the estimated model to check for lack-of-fit. Instead, Tse proposes ``a Lagrange multiplier test for the multinomial logit model against the dogit model as the alternative hypothesis'' (\citeyear{tse_diagnostic_1987}). In general, discrete choice model checks for the ``independence from irrelevant alternatives'' (IIA) property are all model comparisons. Some, as in \citet{tse_diagnostic_1987}, test the MNL model against discrete choice models that do not have the IIA property. Other tests such as the Hausman-McFadden test \citep{hausman_specification_1984}, compare a given MNL model against the same MNL model estimated on a subset of alternatives. Likewise, model comparisons are implicitly or explicitly behind Wald, Lagrange Multiplier, and Likelihood Ratio tests for omitted variables or parameter heterogeneity (both systematic and unobserved heterogeneity as embodied in mixed logit models). All of these tests can be viewed as tests of a restricted model against an unrestricted model.

The major problem with such comparisons is that both models being compared may be misspecified. Accordingly, the fact that one's model passes a model comparison test may not mean anything more than the fact that one's entertained model was better than an even worse model. As noted by \citet{zheng_testing_2008}, ``[...] the performance of these parametric tests rely on correct specification of the alternative model. If the alternative is misspecified, those tests may have low or no power.'' Model checks as described in this paper avoid the issue of mis-specifying one's alternative model by directly testing one's entertained model against the observed data.

\section{Case Study (Part 2)}
\label{sec:case-study-part2}
With a greater understanding of how this paper's proposed methodology is positioned with respect to the relevant literature, this section returns to the case study began earlier. In Section \ref{sec:case1-part1}, I used this paper's proposed model checks to reveal serious underfitting in \citeauthor{brownstone_forecasting_1998}'s MNL model. In this section, I will show how the proposed checks can be used to reduce the observed lack-of-fit.

Overall, the model checking procedure of Section \ref{sec:case1-part1} provided two main insights. First, underfitting was revealed by all of the plots involving the relative price variable (i.e. price over log income). This can be seen, for instance, in Figures \ref{fig:orig-mnl-suv-marginal}, \ref{fig:orig-mnl-electric-kde}, and \ref{fig:orig-mnl-suv-cdf}. To partially address this issue, the relative price variable was re-specified as piecewise linear: one term accounted for relative prices less than 3, and the other term accounted for relative prices greater than 3. Three was chosen as the knot location for the piecewise linear specification based on plots such as Figures \ref{fig:orig-mnl-suv-marginal} and \ref{fig:orig-mnl-suv-cdf}. These graphs showed marked underfitting at the relative price value of 3. For a more formal method of model improvement, one could have instead re-specified the relative price variable using techniques such as generalized additive models \citep{hastie_generalized_1986, abe_generalized_1999}.

Secondly, the ``automatic'' model checking procedure of Section \ref{sec:general-methodology} revealed that for many variables, the pattern of underfitting often differed by both vehicle fuel type and body type. Such differing patterns indicate that there is heterogeneity in how the original MNL model differs from the true data generating process, and as result, there is likely heterogeneity in the true relationship between the various explanatory variables and the probability of a given alternative being chosen. To accommodate such heterogeneity, I re-specified the MNL by including an interaction between the body type variable and the piecewise linear price variable, as well as an interaction between the body type and the range, acceleration, top speed, pollution, vehicle size, and operating cost variables. Similarly, I also added an interaction term between the fuel type variable and the piecewise linear price variables, range, top speed, pollution, vehicle size, and operating cost variables. Acceleration was not interacted with fuel type since the simulated histograms of acceleration by vehicle fuel type did not suggest severe underfitting.

In summary, I changed the specification of \citeauthor{brownstone_forecasting_1998}'s MNL model to include a piecewise linear specification of the relative price variable and to include interactions between many of the explanatory variables and the vehicle body type and fuel type. These changes were not exhaustive of all improvements suggested by my proposed model checking procedure. For instance, simulated histograms of the range, top speed, pollution, vehicle size and operating cost variables all suggest that these variables should not be entered using a single linear term in the model. These variables have few unique values, and given that the patterns of underfitting vary across these values, the variables in question should be represented as categorical (e.g. using dummy-variables) rather than continuous. Despite recognizing these issues, I restricted myself to the changes described in this paragraph's opening sentence for simplicity\footnote{Using dummy variables for the vehicle attributes mentioned above would have necessitated the use of regularization to ensure monotonicity in the (ordered) attribute values.}. In particular, my aim in this section is to show that the proposed model checking procedures can suggest concrete model improvements, not to demonstrate how one can make \textit{all possible} improvements and not to demonstrate the best possible way to make model improvements.

As a result, the parameter estimates are not of major importance in this study. For space reasons, interested readers will therefore have to find the typical table of parameter estimates in Appendix B.

\begin{table}
\centering
\begin{tabular}{lrr}
\toprule
{Summary Measure}                   &           Original    &    Expanded \\
\midrule
In-sample Log-Likelihood            &      -7,391.830    &    -7,311.634 \\
McFadden's $\bar{\rho}^2$         &               0.111    &            0.113 \\
AIC 										   &            14,825     &          14,787 \\
Out-of-sample Log-Likelihood     &         -741.183 &           -739.723 \\

\bottomrule
\end{tabular}
\caption{Model Summary Statistics}
\label{table:model-summaries}
\end{table}

Instead, Table \ref{table:model-summaries} provides general summary statistics about the fits of the original and expanded MNL models. This table's main message is that there are multiple points of corroborating, classical evidence that the additional parameters in the expanded MNL are useful and not simply overfitting. With all the changes described two paragraphs ago, the number of estimated parameters increased from 21 coefficients in \citeauthor{brownstone_forecasting_1998}'s MNL model to 89 in the expanded model. Accordingly, the in-sample log-likelihood of the expanded model increased to -7,311.634---higher than that of the MNL, probit, or mixed logit models estimated in \citet{brownstone_forecasting_1998}. Even after penalizing the expanded model for its extra parameters, McFadden's $\bar{\rho}^2$ was higher in the expanded MNL (0.113) versus the original MNL model (0.111). Likewise, the Akaike Information Criterion (AIC) was lower in the expanded MNL (14,787) versus the original MNL (14,825). Finally, the out-of-sample log-likelihood\footnote{These values were computed using the average out-of-sample log-likelihood from ten-fold cross-validation. } was greater for the expanded MNL model (-739.7) as compared to the original model (-741.2).

\begin{figure}
\centering
\includegraphics[width=0.6\textwidth]{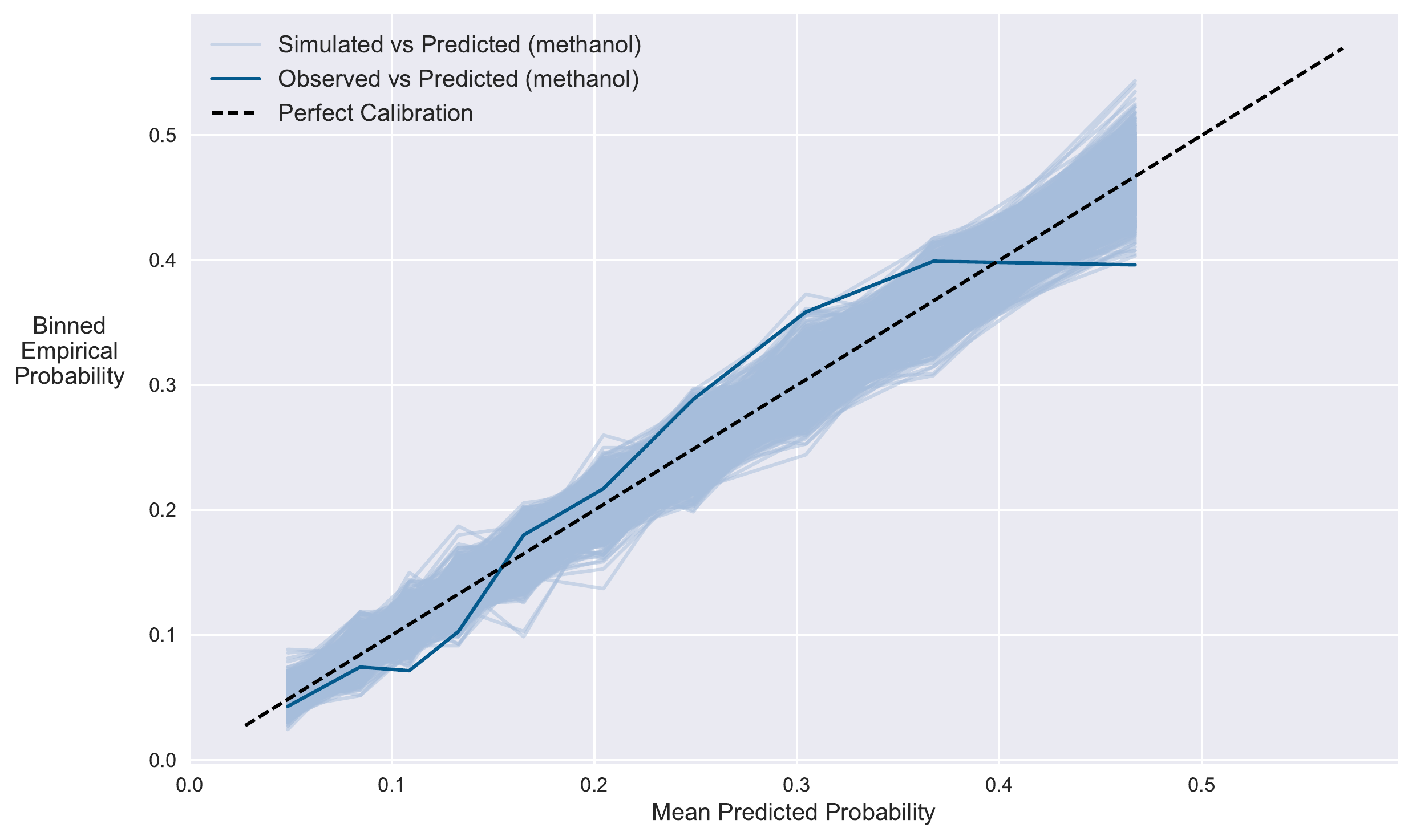}
\caption{Binned reliability plot for methanol vehicles using the `expanded' MNL model}
\label{fig:new-mnl-methanol-reliability}
\end{figure}

\begin{figure}
\centering
\includegraphics[width=0.6\textwidth]{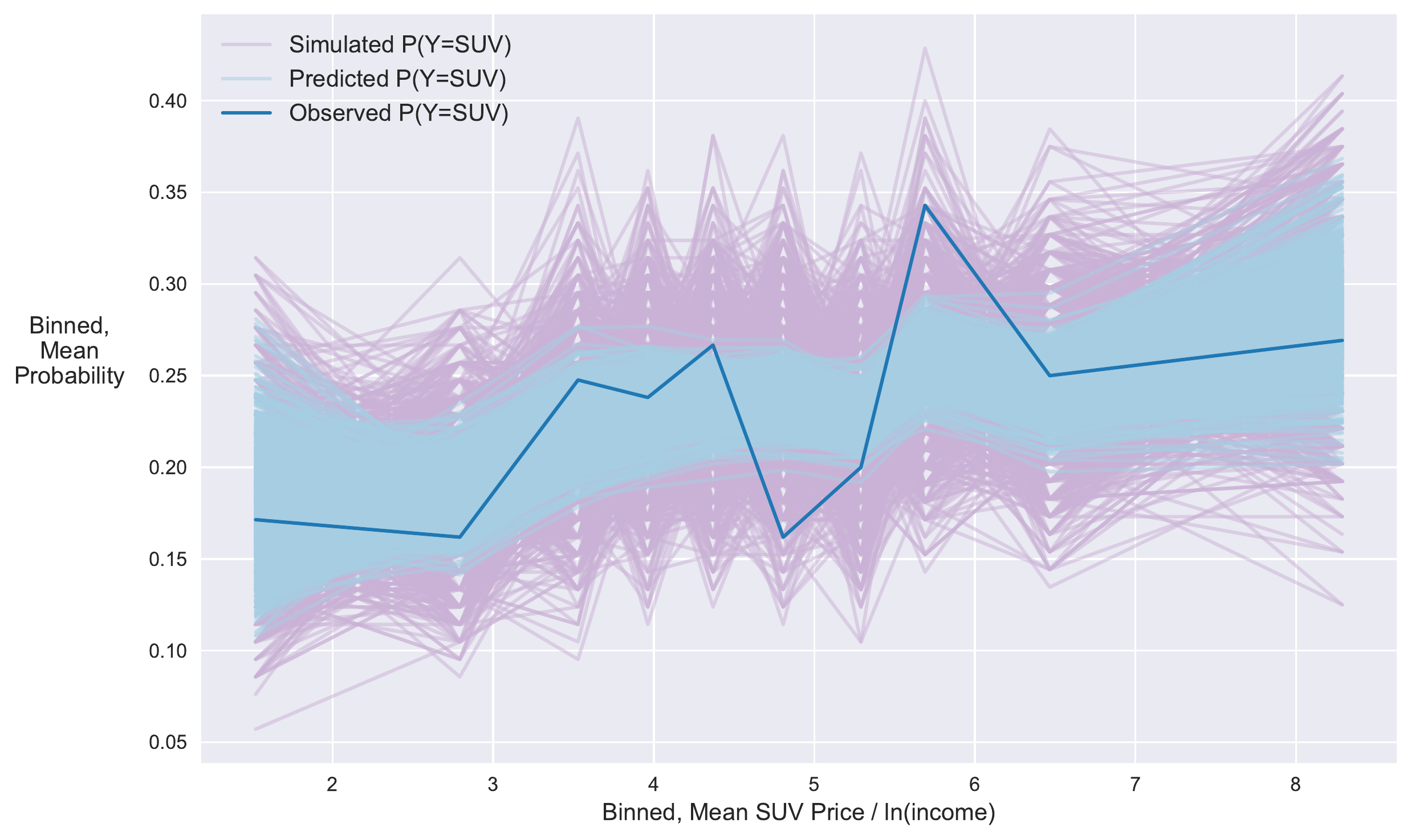}
\caption{Binned marginal model plot for sports utility vehicles using the `expanded' MNL model}
\label{fig:new-mnl-suv-marginal}
\end{figure}

\begin{figure}
\centering
\includegraphics[width=0.6\textwidth]{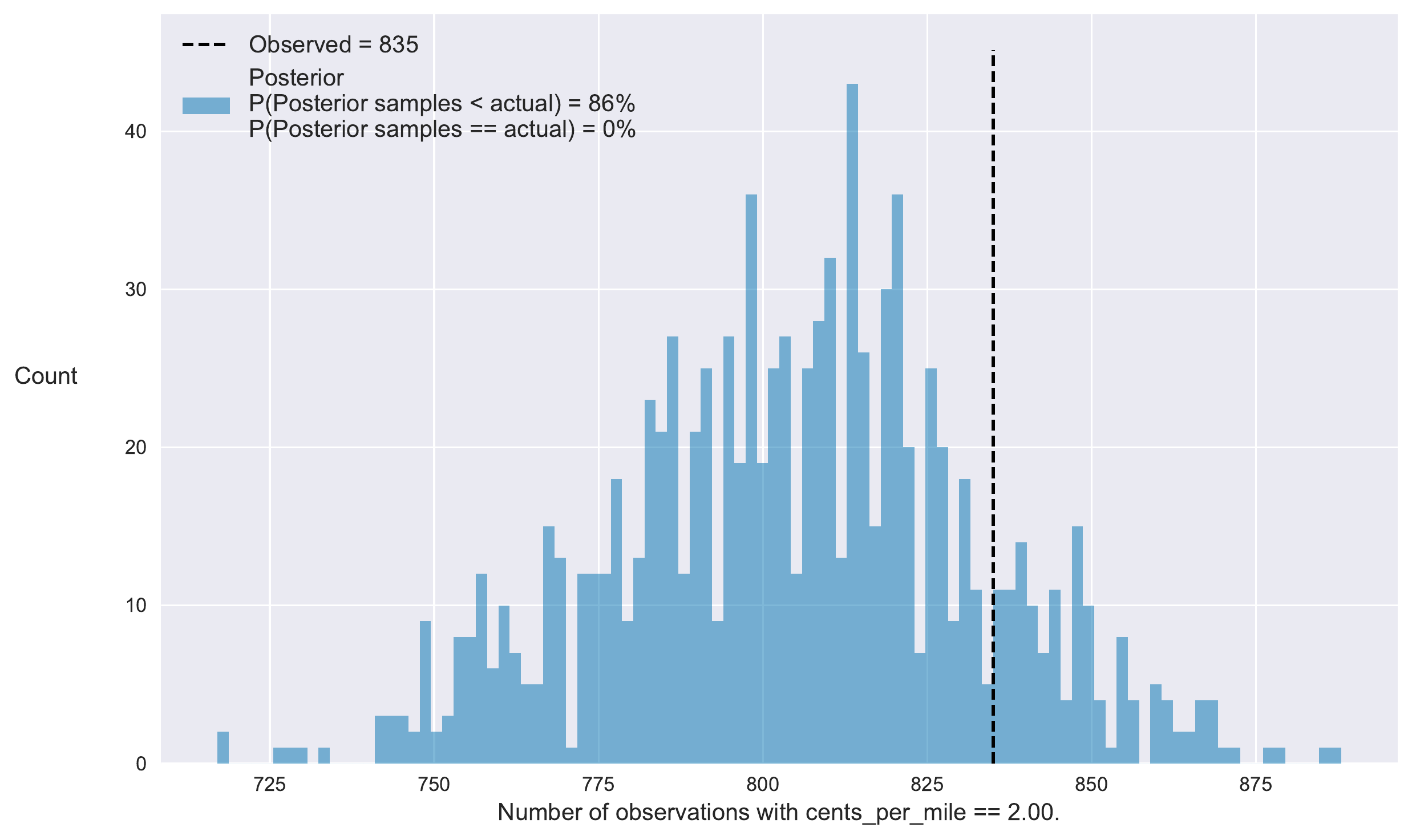}
\caption{Simulated histograms for regular cars using the `expanded' MNL model}
\label{fig:new-mnl-regcar-histogram}
\end{figure}

\begin{figure}
\centering
\includegraphics[width=0.6\textwidth]{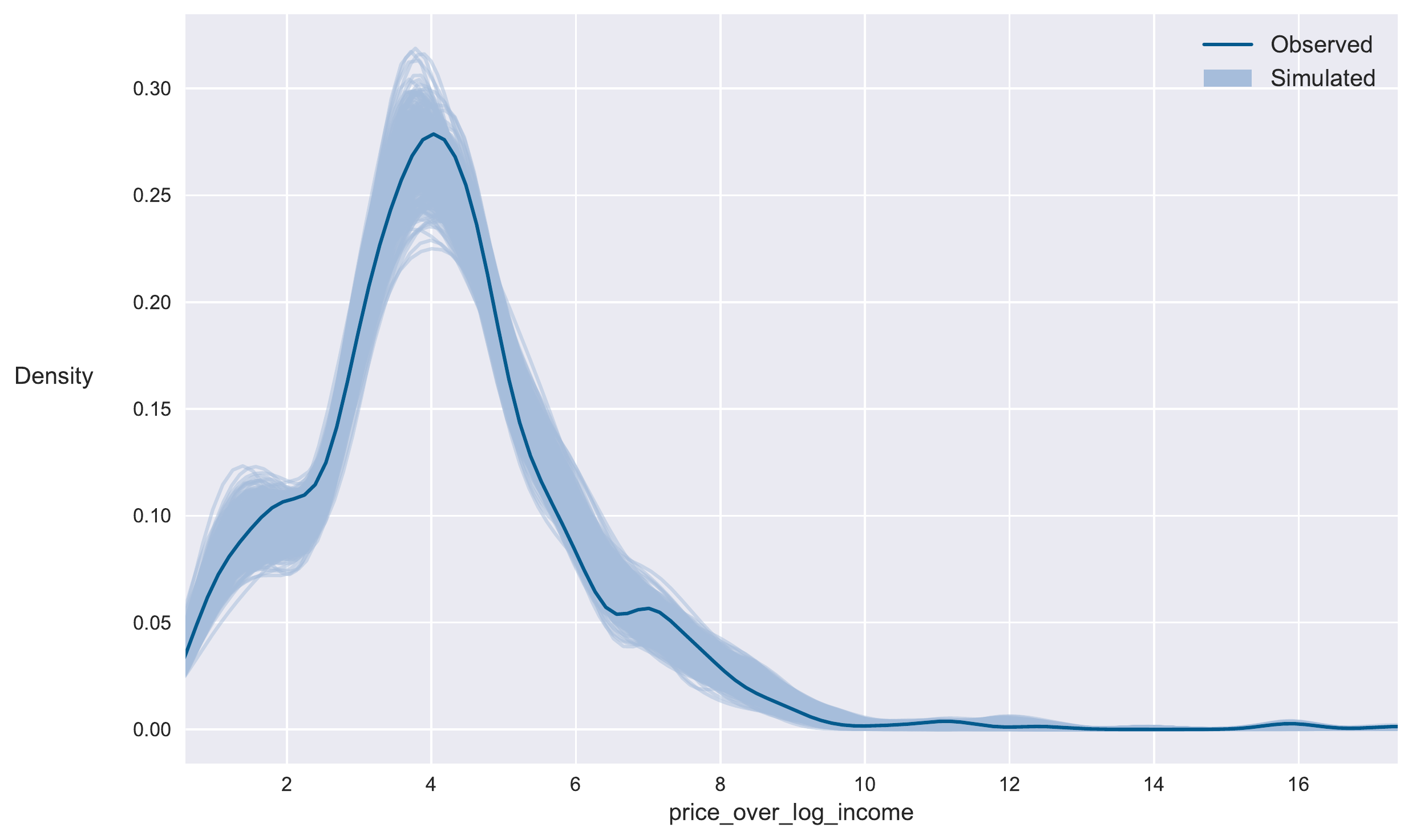}
\caption{Simulated kernel density estimates for electric cars using the `expanded' MNL model}
\label{fig:new-mnl-electric-kde}
\end{figure}

\begin{figure}
\centering
\includegraphics[width=0.6\textwidth]{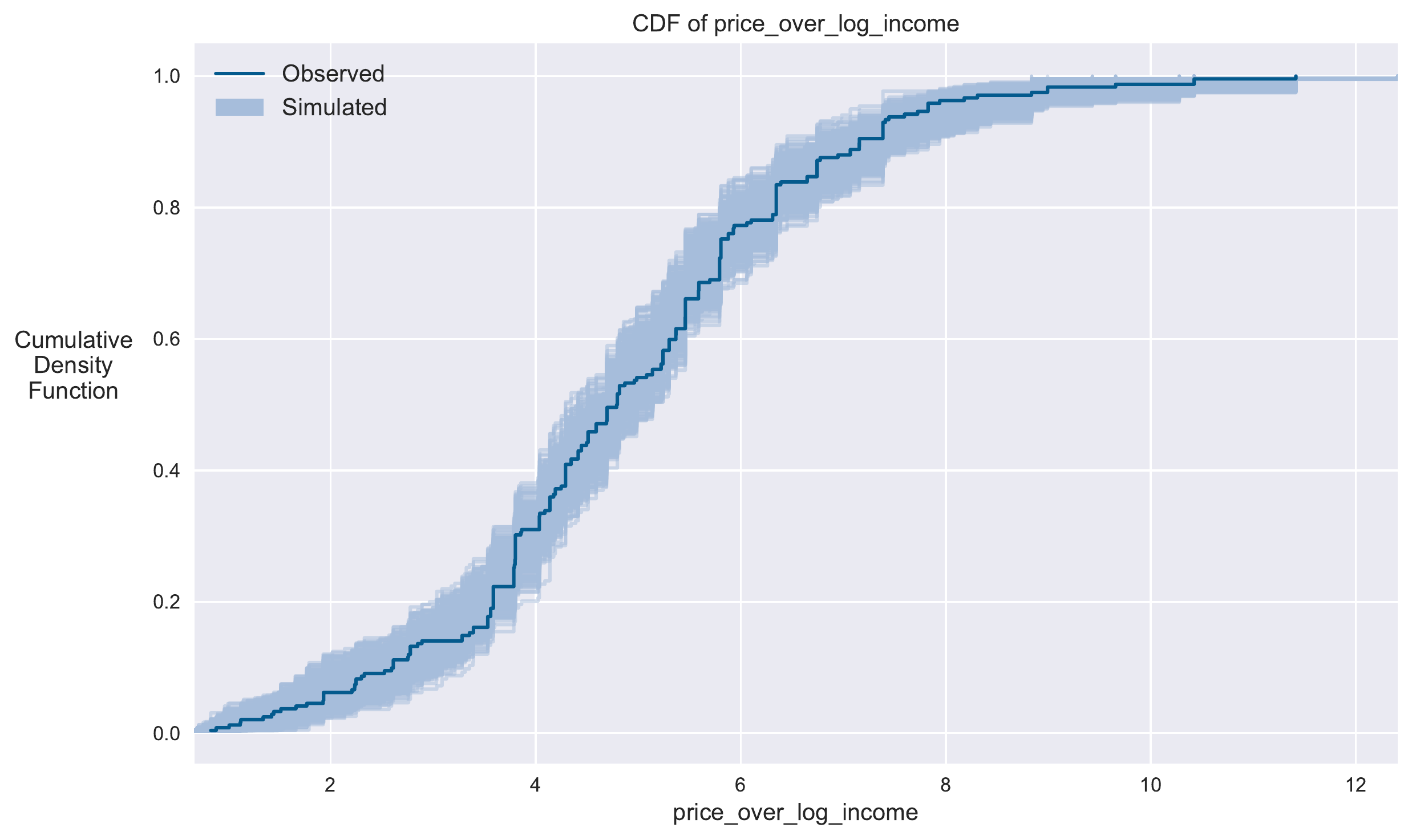}
\caption{Simulated CDFs for sport utility vehicles using the `expanded' MNL model}
\label{fig:new-mnl-suv-cdf}
\end{figure}

Beyond classical measures of fit, the expanded MNL model was subjected to the same graphical model checking procedures as the original model. As expected, and as shown in Figures \ref{fig:new-mnl-methanol-reliability} - \ref{fig:new-mnl-suv-cdf}, many of the observed instances of underfitting from the original MNL model have been resolved. In particular, Figures \ref{fig:new-mnl-suv-marginal}, \ref{fig:new-mnl-electric-kde}, and \ref{fig:new-mnl-suv-cdf} show that the underfitting of the relative price variable has been greatly reduced. For instance, the observed KDE no longer lies outside the simulated KDEs at any point, and the observed CDF now lies in the middle of the simulated CDFs. Likewise, the marginal model plots have improved: in the expanded model, only two points lie outside the blue band of asymptotically expected probabilities, whereas 7 points lied outside this band in the \citeauthor{brownstone_forecasting_1998}'s MNL model. In terms of the operating cost, Figure \ref{fig:new-mnl-regcar-histogram} shows that the observed data is less extreme in the expanded MNL model as compared to the original: 86\% of predicted samples in the expanded model are less than the observed value as compared to 96\% with the original MNL model. Finally, by comparing Figure \ref{fig:new-mnl-methanol-reliability} to Figure \ref{fig:orig-mnl-methanol-reliability}, one can see that the expanded MNL model has better calibrated (but still unreliable) probability predictions. For methanol vehicles, the area between the dark blue curve and the dashed reference line is lower in Figure \ref{fig:new-mnl-methanol-reliability} with the expanded MNL model as compared to Figure \ref{fig:orig-mnl-methanol-reliability} with \citeauthor{brownstone_forecasting_1998}'s MNL model. This finding is also confirmed by the reliability plots (not shown due to space considerations) for the other fuel types.

Finally, the changes inspired by this paper's techniques do more than merely improve the fit of the original MNL model. The changes also lead to substantively different policy implications when compared to \citeauthor{brownstone_forecasting_1998}'s chosen mixed logit model (``Mixed logit B''). In particular, one main policy that was analyzed in \citet{brownstone_forecasting_1998} was a 20\% increase in the price of large gasoline cars. \citeauthor{brownstone_forecasting_1998}'s chosen model predicted that the market share of large gas cars would have a relative decrease of approximately 12\% percent\footnote{Forecast results for \citeauthor{brownstone_forecasting_1998}'s Mixed logit B were not numerically reported in their paper. The forecast results in this text were manually produced using custom written python code and \citeauthor{brownstone_forecasting_1998}'s estimated parameters.}. Similarly, the expanded MNL model predicted an 8\% relative decrease in the market share of large gas cars.

\begin{table}
\centering
\begin{tabularx}{0.75\textwidth}{QQ}
\toprule
Mixed logit B (\%)                &           Expanded MNL (\%)    \\

\midrule

Large gasoline truck \newline (8.16)  &  Sub-compact electric van \newline (3.03) \\
Large gasoline station wagon \newline (6.15)  &  Compact CNG station wagon \newline (3.02)  \\
Large gasoline van \newline (5.72)  &  Large gasoline truck \newline (2.88) \\
Compact CNG station wagon \newline (4.14)  &  Large gasoline station wagon \newline (2.42) \\

\bottomrule
\end{tabularx}
\caption{Top-Four Vehicle Categories Forecasted to Increase in Market Share}
\label{table:vehicle-forecasts}
\end{table}

Despite similar predictions for the market share of large gas cars, the two model's forecasts differ most regarding the vehicle classes that are predicted to grow appreciably in market share. These differences are shown in Table \ref{table:vehicle-forecasts}. On one hand, \citeauthor{brownstone_forecasting_1998}'s chosen model predicts an environmentally ``hellish'' scenario where people buy other, even larger gasoline powered vehicles. On the other hand, my expanded MNL model predicts an environmentally ``heavenly'' scenario where individuals buy smaller, electric and compressed natural gas vehicles.

Specifically, \citeauthor{brownstone_forecasting_1998}'s chosen model predicts that large gasoline station wagons, trucks, and vans would see relative increases in their market share of approximately 6\%, 8\%, and 6\% respectively. Moreover, \citeauthor{brownstone_forecasting_1998}'s chosen model does not predict that any alternative fuel vehicle will have a relative market share increase greater than approximately 4\%. In contrast, my expanded MNL model predicts that subcompact electric vans and compact, compressed natural gas station wagons will have the largest relative market share increases, at 3\% each. After this, the expanded MNL model predicts relative market share increases of 2\% for both large gasoline vehicles as well as many alternative fuel vehicles. These differing forecasts between Mixed logit B and the expanded MNL model imply conflicting qualitative lessons for policy makers that may have been considering policies such as environmentally motivated taxes on large gasoline cars.

\section{Conclusion}
\label{sec:conclusion}
In this paper, I introduced and demonstrated a procedure for checking one's discrete choice models using predictive simulations. Conceptually, this method compares features of one's observed data to the distribution of those features in the datasets formed using simulated outcomes. To operationalize these ideas, I introduced seven graphical model checks for underfitting. These seven plots are designed to be of use in routine model checking situations, ideally throughout and at the conclusion of any discrete choice analysis. In addition to these specific plots, I also propose a general and nearly-automatic procedure for designing custom checks for underfitting. To demonstrate the proposed procedures, I used a vehicle choice case study from \citet{brownstone_forecasting_1998}. In this case study, the seven plots and the proposed model checking procedure uncovered serious underfitting in \citeauthor{brownstone_forecasting_1998}'s MNL model---underfitting that was not addressed by their final mixed logit model. I then demonstrated how the insights derived from my proposed plots lead to an expanded MNL model with greater levels of fit and substantively different forecasts from \citeauthor{brownstone_forecasting_1998}'s final mixed logit model. In particular, this paper's proposed model checks inspired a more detailed representation of the systematic heterogeneity in individual sensitivities to variables such as price, operating cost, etc. Once this systematic heterogeneity was better captured, the resulting model fit better than either \citeauthor{brownstone_forecasting_1998}'s MNL or mixed logit models. Moreover, the new model had aggregate forecasts that were much more hopeful about the ability of pricing policies to push individuals away from large gas cars to smaller, alternative fuel vehicles.

Beyond this specific example, the goal of this paper is to introduce a new set of model checking tools to the discrete choice community. The case study of \citet{brownstone_forecasting_1998} was used to show how models estimated by the best of us may still be underfitting our data. In my own work, I have used these techniques to spot errors even more egregious than those detailed in this paper. These errors have been found everywhere: from simple binary logit models to complex machine learning techniques. No model is safe. However, with the model checking methods introduced in this paper, we can better see how our models are misrepresenting reality, thereby allowing us to raise the quality of models used within our papers, our companies, and our governments.

\section*{Acknowledgements}
Thanks to Albert Yuen, Joan Walker, Andrew Campbell, and two reviewers for providing helpful comments on an earlier draft of this manuscript. Any remaining errors are my own. Thanks also to Andrew Campbell for inspiring this research by asking the fundamental question: ``so, how do you check your models after you've estimated them?''

\section*{Replication Materials}
All replication data and code for this paper can be found online at \url{https://github.com/timothyb0912/check-yourself}.

Additionally, stand-alone software for producing the posterior predictive simulations and plots mentioned in this paper can be found in the ``pylogit.viz'' module of the PyLogit open-source package for the python programming language. This software will work regardless of the software that one initially used for model estimation. One simply needs to be able to store one's data and simulated probability predictions in a software-agnostic format such as a csv file. The package is available through the Python Packaging Index (PyPI) as well as through the Anaconda distribution. See \url{https://github.com/timothyb0912/pylogit} for detailed installation instructions.

\newpage
\begin{appendices}

\section{Frequently Asked Questions}
\label{sec:freq-questions}
This paper introduced the concept of assessing underfitting in discrete choice models by visually comparing predictive simulations from one's model to one's observed data. This introduction has included concrete examples of such graphical model checks in Section \ref{sec:case1-part1}, a general algorithm for use with labeled and unlabeled alternatives in Section \ref{sec:general-methodology}, relations to the broader literature on model checking in statistics and discrete choice in Section \ref{sec:lit-review}, and examples of how these model checks could be used to guide one's model improvement efforts in Section \ref{sec:case-study-part2}.

However, despite all the topics discussed in this introduction, some related issues have still been excluded in order to minimize the article's length. This appendix on \textit{Frequently Asked Questions} attempts to briefly address the most common questions that reviewers, friends, and colleagues have had after reading earlier drafts of this paper.

\begin{QandA}
   \item How do predictive simulations help us make decisions about how to improve our models?
       \begin{answered}
           Essentially, predictive simulations are like thermometers. Thermometers are useful for assessing a patient's body temperature, and through repeated use, one can assess how a patient's temperature is changing alongside a caretaker's actions. Likewise, predictive simulations are most useful for assessing the discord between one's model and reality, and through repeated use, one can assess how this discord has been altered by changes to one's model. Taking the analogy further, note that thermometers do not tell doctors the cause of a fever or how a fever can be reduced. Similarly, predictive simulations do not tell researchers the cause of the problems they uncover, nor do they suggest particular remedies to those problems. Once aware of the extent to which one's current model misrepresents reality, it is the analyst's responsibility to diagnose their model and brainstorm ways to increase its realism. Afterwards, predictive simulations will be there to show the analyst how well their changes worked.
       \end{answered}
       
   \begin{figure}
    \centering
	\includegraphics[width=0.6\textwidth]{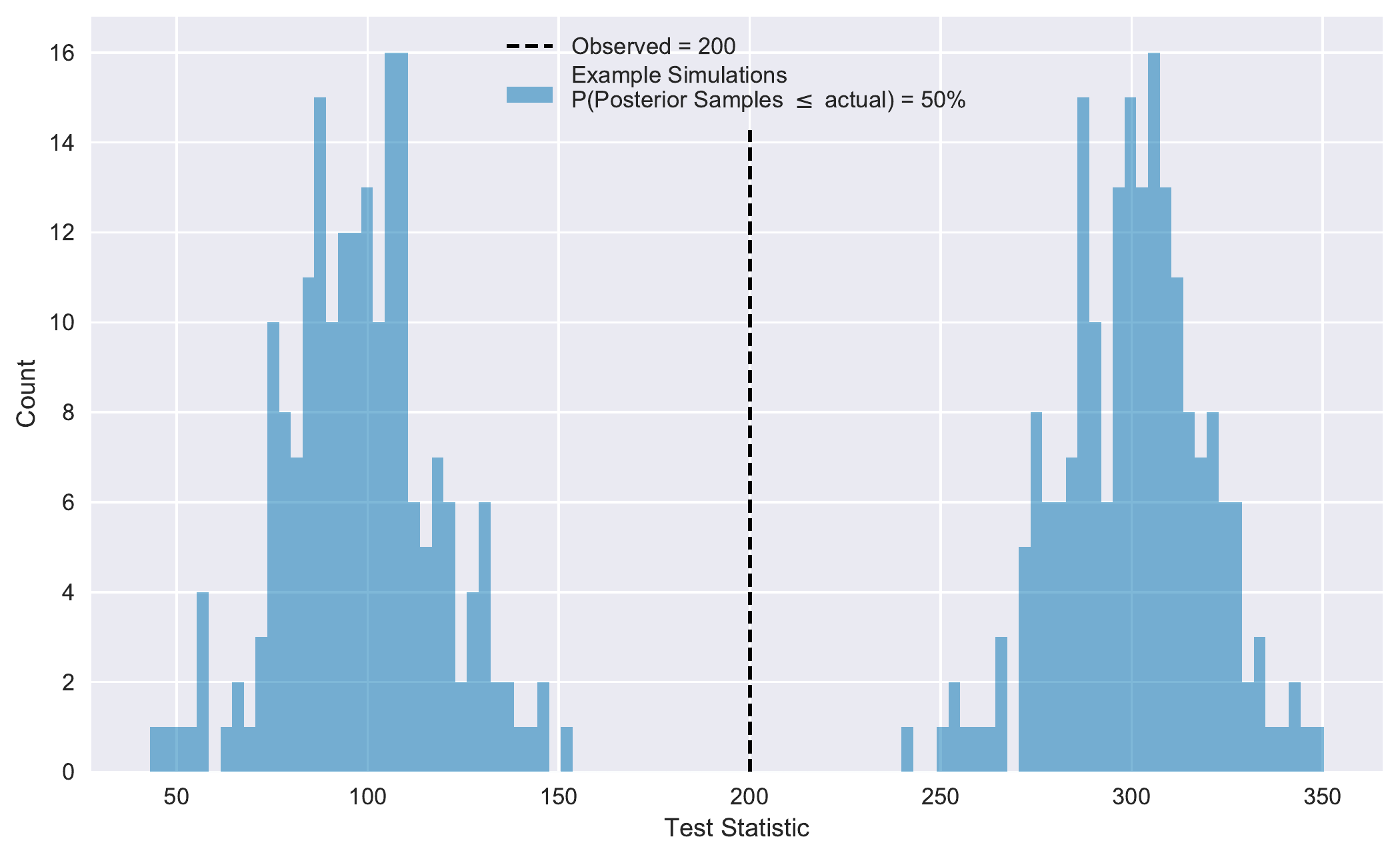}
	\caption{Example histogram where the predictive p-value is misleading}
	\label{fig:example-histogram}
	\end{figure}
   
   \item Why should we use graphical model checks instead of simply looking at point measures or purely numeric summaries?
       \begin{answered}
           In general, solely looking at scalar values (such as p-values) leads to a loss of information. Far from being harmless, this information loss can potentially lead to incorrect assessments of model fit. Imagine one is looking at a simulated histogram such as the one shown in Figure \ref{fig:example-histogram}. Here, one's predictive distribution is bimodal and is roughly symmetric about the observed value. As a result, one's p-value may be close to 0.5, even though few simulated datasets in this example have test statistics that are similar to the observed test statistic (the dashed vertical line in the plot). To facilitate correct assessments of model fit and avoid being unnecessarily misled, I recommend using graphical model checks in addition to purely numeric summaries.
           
           Beyond the insufficiency of predictive p-values for model checking, creating other scalar surrogates for the graphical check is non-trivial. For instance, even if one's predictive distribution is unimodal, it is not valid to think that the value of a test statistic, evaluated using the probabilities generated at a point estimate of one's parameters, will serve as a useful surrogate for the mean of one's predictive distribution of that test statistic. This is because $ E \left[ f \left( z \right) \right] \neq f \left( E \left[ z \right] \right)$ with respect to a random variable $z$ and arbitrary non-linear function $f \left( \cdot \right)$. In this paper's model checking context, this fact implies that even though $\beta_{\textrm{MLE}}$ is the mean of the asymptotic sampling distribution, and therefore the mean of the simulated parameters $\beta_r$, $P \left( Y \mid X, \beta_{\textrm{MLE}} \right) \neq E \left[ P \left( Y \mid X, \beta_r \right) \right]$. The probabilities at our point estimate of the parameters are not the mean of the simulated probabilities. Given this inequality, even simple graphical checks such as simulated histograms cannot be replaced by scalar comparisons based on point estimates. More complicated graphical checks such as simulated KDEs use an additional layer of non-linear functions in their calculation of the test statistic. These additional non-linearities rule out the use of a point estimate of the model's parameters for purposes of trivially computing the mean of the predictive distributions of one's test statistics.
       \end{answered}
   
   \item What are some ways that predictive simulations should NOT be used?
       \begin{answered}
           Predictive simulations should not be used as hypothesis tests that are used to accept or reject models. A-priori, we know all models are wrong. Predictive simulations should therefore be used to understand the ways and extent to which one's model differs from reality. If desired, this understanding can then be used increase the realism of one's model or to inform the way that the model is used. In some cases, this informing may mean completely rejecting the use of a particular model. However, in other cases, one may decide to only make inferences from the model when the model performs well on checks related to the inference in question. The decision is up to the ``stakeholders'' of the model (e.g. peer reviewers, company or organization managers, etc.). For example, if a travel mode choice model performs well on underfitting checks related to transit travel cost but not travel time, then perhaps one may use the model to forecast the impact of fare changes but not for forecasting the impact of adding an express service that would reduce individual travel times.
       \end{answered}

   \item How do we interpret the proposed graphical model checks?
       \begin{answered}
           In general, the proposed graphical model checks are ways to visually assess the similarity between data generated by one's model and one's observed data. As mentioned in the question ``How do predictive simulations help us make decisions about how to improve our models?,'' the proposed checks are to be used like a thermometer. One checks one's temperature, and any suspiciously high temperature measurements must then be investigated for their root cause and effective remedies. Similarly, any discrepancies uncovered by the proposed model checks have to be investigated for their root cause and for appropriate model improvements. Such investigations may not be easy or straightforward. The visually diagnosed problems may be related to one's overall model specification (e.g. using an MNL model instead of a nested logit model), to omitted variables, or even to misspecification of the relationship between one or more correlated variables and the outcome variable. As usual, an analyst must still make use of their domain knowledge, intuition, and thoroughness to resolve any discovered problems.
           Similar to the question of interpreting graphical model checks is the question of how one interprets their associated predictive p-values. Mathematically, the predictive p-value is the proportion of simulated datasets for which a given scalar statistic was lower than the observed value of that scalar statistic. These predictive p-values can be helpful (though insufficient, in my opinion) for evaluating the extent to which one's observed data ``looks like'' one's simulated data. In particular, they may be especially helpful for assessing small model changes that are hard to distinguish visually in a graphical model check.
       \end{answered}

   \item When should one be concerned by the results of a particular graphical model check?
       \begin{answered}
           I do not think that there is a universal answer to this question. It is equivalent to asking ``how good is good enough?'' One's answer to these questions will almost assuredly be driven by problem specific considerations such as the purpose for which the model is being built and the specific statistic being checked. However, there are some general guidelines that one can use. Overall, if one's model is close to the data generating process, then one's observed data should look like a random draw of the data simulated from one's model. Accordingly, one should expect the observed statistics to lie near the center of the distribution of simulated statistics. For scalar statistics, I would be concerned if the distribution of simulated statistics was not clustered around a line for the observed statistic. For vector statistics, I would be concerned if the observed vector, at all points, was not near the center of the simulated vectors. Additionally I would be concerned if the observed vector did not follow the same visual trends as the simulated vectors. (After all, the observed vector should ``look like'' a random draw of the simulated vectors). How far away from this ideal a model is allowed to exist is up to the ``stakeholders'' of the model to decide and to defend.
       \end{answered}

   \item Besides the examples shown in this paper, what are other specific ways that one can use predictive simulations to check our models?
       \begin{answered}
           What follows is by no means an exhaustive list, but predictive simulations can also be used in the following four ways to check one's models. First, as demonstrated in \citet{pardoe_2002_graphical}, one can use predictive simulations to determine whether variables should be added or removed from one's model. Secondly, one can use the proposed graphical checks to visualize and understand differences between competing model extensions. For instance, one can check the data generated from using a common specification of one's systematic utilities but different distributions of the random utility (i.e. different link functions), different mixing distributions in a mixed logit model, different class membership models in a latent class model, etc. Third, a closely related use of predictive simulations is to check one's mixed logit models by comparing one's inferences for the distribution individual tastes with one's assumed mixing distribution of tastes (see \citet{allenby_marketing_1998} and \citet{gilbride_posterior_2010} for examples). Finally, as demonstrated and explained in \citet{kucukelbir_evaluating_2017}, one could also use predictive simulations to discover groups of observations that one's model fits particularly poorly.
       \end{answered}
\end{QandA}

\newpage
\section{Expanded MNL Estimation Results}
\label{sec:appendix-b-expanded-results}
\begin{table}[!htbp]
\centering
\begin{tabular}{lrrrr}
\toprule
\textbf{Variable} & \textbf{Estimate} & \textbf{Std. err} & \textbf{z} & \textbf{P > |z| }  \\
\midrule
\textbf{Price over log(income) <= 3 (sports\_utility\_vehicle)}  &       0.2328  &        0.257     &     0.905  &         0.366        \\
\textbf{Price over log(income) <= 3 (sports\_car)}               &      -0.1069  &        0.253     &    -0.423  &         0.673        \\
\textbf{Price over log(income) <= 3 (station\_wagon)}            &      -0.3201  &        0.109     &    -2.937  &         0.003        \\
\textbf{Price over log(income) <= 3 (truck)}                     &      -0.2541  &        0.089     &    -2.863  &         0.004        \\
\textbf{Price over log(income) <= 3 (van)}                       &      -0.2418  &        0.088     &    -2.741  &         0.006        \\
\textbf{Price over log(income) <= 3 (electric)}                  &      -0.1319  &        0.100     &    -1.324  &         0.186        \\
\textbf{Price over log(income) <= 3 (compressed\_natural\_gas)}  &      -0.0938  &        0.092     &    -1.024  &         0.306        \\
\textbf{Price over log(income) <= 3 (methanol)}                  &       0.0629  &        0.089     &     0.707  &         0.480        \\
\textbf{Price over log(income) <= 3}                             &      -0.2519  &        0.113     &    -2.225  &         0.026        \\
\textbf{Price over log(income) $>$ 3 (sports\_utility\_vehicle)} &       0.2714  &        0.077     &     3.508  &         0.000        \\
\textbf{Price over log(income) $>$ 3 (sports\_car)}              &       0.2039  &        0.080     &     2.534  &         0.011        \\
\textbf{Price over log(income) $>$ 3 (station\_wagon)}           &      -0.0341  &        0.047     &    -0.722  &         0.470        \\
\textbf{Price over log(income) $>$ 3 (truck)}                    &      -0.0087  &        0.035     &    -0.249  &         0.804        \\
\textbf{Price over log(income) $>$ 3 (van)}                      &      -0.0400  &        0.036     &    -1.110  &         0.267        \\
\textbf{Price over log(income) $>$ 3 (electric)}                 &      -0.0941  &        0.035     &    -2.655  &         0.008        \\
\textbf{Price over log(income) $>$ 3 (compressed\_natural\_gas)} &      -0.0534  &        0.033     &    -1.616  &         0.106        \\
\textbf{Price over log(income) $>$ 3 (methanol)}                 &      -0.0699  &        0.031     &    -2.253  &         0.024        \\
\textbf{Price over log(income) $>$ 3}                            &      -0.1326  &        0.037     &    -3.553  &         0.000        \\
\textbf{Range (units: 100mi) (sports\_utility\_vehicle)}         &      -0.0964  &        0.100     &    -0.965  &         0.334        \\
\textbf{Range (units: 100mi) (sports\_car)}                      &      -0.0721  &        0.117     &    -0.616  &         0.538        \\
\textbf{Range (units: 100mi) (station\_wagon)}                   &      -0.0725  &        0.086     &    -0.848  &         0.397        \\
\textbf{Range (units: 100mi) (truck)}                            &       0.0125  &        0.070     &     0.178  &         0.859        \\
\textbf{Range (units: 100mi) (van)}                              &       0.0628  &        0.067     &     0.939  &         0.348        \\
\textbf{Range (units: 100mi) (electric)}                         &       0.1430  &        0.130     &     1.098  &         0.272        \\
\textbf{Range (units: 100mi) (compressed\_natural\_gas)}         &       0.2916  &        0.132     &     2.203  &         0.028        \\
\textbf{Range (units: 100mi) (methanol)}                         &       0.3219  &        0.135     &     2.393  &         0.017        \\
\textbf{Range (units: 100mi)}                                    &       0.1028  &        0.115     &     0.893  &         0.372        \\
\textbf{Acceleration (units: 0.1sec) (sports\_utility\_vehicle)} &      -0.1927  &        0.497     &    -0.388  &         0.698        \\
\textbf{Acceleration (units: 0.1sec) (sports\_car)}              &      -1.0425  &        0.602     &    -1.732  &         0.083        \\
\textbf{Acceleration (units: 0.1sec) (station\_wagon)}           &      -0.3109  &        0.446     &    -0.698  &         0.485        \\
\textbf{Acceleration (units: 0.1sec) (truck)}                    &      -0.1371  &        0.360     &    -0.381  &         0.704        \\
\textbf{Acceleration (units: 0.1sec) (van)}                      &      -0.6375  &        0.323     &    -1.977  &         0.048        \\
\textbf{Acceleration (units: 0.1sec)}                            &      -0.5157  &        0.146     &    -3.538  &         0.000        \\
\textbf{Top speed (units: 0.01mph) (sports\_utility\_vehicle)}   &      -0.0701  &        0.331     &    -0.212  &         0.832        \\
\textbf{Top speed (units: 0.01mph) (sports\_car)}                &       0.2942  &        0.399     &     0.737  &         0.461        \\
\textbf{Top speed (units: 0.01mph) (station\_wagon)}             &       0.1550  &        0.282     &     0.550  &         0.582        \\
\textbf{Top speed (units: 0.01mph) (truck)}                      &       0.1298  &        0.224     &     0.579  &         0.562        \\
\textbf{Top speed (units: 0.01mph) (van)}                        &      -0.0136  &        0.218     &    -0.063  &         0.950        \\
\textbf{Top speed (units: 0.01mph) (electric)}                   &       0.5526  &        0.325     &     1.698  &         0.089        \\
\bottomrule
\end{tabular}

\caption{Expanded MNL model}
\label{table:vehicle-choice-expanded-mnl-results}
\end{table}

\begin{table}
\centering
\begin{tabular}{lrrrr}
\toprule
\textbf{Variable} & \textbf{Estimate} & \textbf{Std. err} & \textbf{z} & \textbf{P > |z| }  \\
\midrule

\textbf{Top speed (units: 0.01mph) (compressed\_natural\_gas)}   &       0.0994  &        0.286     &     0.348  &         0.728        \\
\textbf{Top speed (units: 0.01mph) (methanol)}                   &      -0.1352  &        0.269     &    -0.503  &         0.615        \\
\textbf{Top speed (units: 0.01mph)}                              &       0.1290  &        0.194     &     0.665  &         0.506        \\
\textbf{Pollution (sports\_utility\_vehicle)}                    &       0.0771  &        0.332     &     0.232  &         0.816        \\
\textbf{Pollution (sports\_car)}                                 &       0.2570  &        0.376     &     0.683  &         0.495        \\
\textbf{Pollution (station\_wagon)}                              &      -0.5810  &        0.280     &    -2.077  &         0.038        \\
\textbf{Pollution (truck)}                                       &       0.0841  &        0.214     &     0.394  &         0.694        \\
\textbf{Pollution (van)}                                         &      -0.1365  &        0.216     &    -0.633  &         0.527        \\
\textbf{Pollution (compressed\_natural\_gas)}                    &       0.0576  &        0.314     &     0.184  &         0.854        \\
\textbf{Pollution (methanol)}                                    &       0.0379  &        0.266     &     0.143  &         0.887        \\
\textbf{Pollution}                                               &      -0.4618  &        0.164     &    -2.813  &         0.005        \\
\textbf{Size (sports\_utility\_vehicle)}                         &       2.4675  &        0.943     &     2.617  &         0.009        \\
\textbf{Size (sports\_car)}                                      &      -2.5383  &        0.988     &    -2.569  &         0.010        \\
\textbf{Size (station\_wagon)}                                   &       0.9575  &        0.851     &     1.126  &         0.260        \\
\textbf{Size (truck)}                                            &      -0.2520  &        0.719     &    -0.351  &         0.726        \\
\textbf{Size (van)}                                              &       2.7231  &        0.825     &     3.299  &         0.001        \\
\textbf{Size (electric)}                                         &      -0.2671  &        0.769     &    -0.347  &         0.728        \\
\textbf{Size (compressed\_natural\_gas)}                         &      -0.4512  &        0.759     &    -0.595  &         0.552        \\
\textbf{Size (methanol)}                                         &       0.4433  &        0.721     &     0.615  &         0.539        \\
\textbf{Size}                                                    &       0.7798  &        0.616     &     1.266  &         0.205        \\
\textbf{Big enough}                                              &       0.1208  &        0.078     &     1.543  &         0.123        \\
\textbf{Luggage space}                                           &       0.4516  &        0.194     &     2.324  &         0.020        \\
\textbf{Operation cost (sports\_utility\_vehicle)}               &       0.5759  &        0.313     &     1.842  &         0.065        \\
\textbf{Operation cost (sports\_car)}                            &       0.5041  &        0.377     &     1.338  &         0.181        \\
\textbf{Operation cost (station\_wagon)}                         &      -0.5383  &        0.280     &    -1.926  &         0.054        \\
\textbf{Operation cost (truck)}                                  &       0.3537  &        0.212     &     1.672  &         0.095        \\
\textbf{Operation cost (van)}                                    &       0.0033  &        0.205     &     0.016  &         0.987        \\
\textbf{Operation cost (electric)}                               &       0.2615  &        0.231     &     1.133  &         0.257        \\
\textbf{Operation cost (compressed\_natural\_gas)}               &      -0.1772  &        0.273     &    -0.648  &         0.517        \\
\textbf{Operation cost (methanol)}                               &      -0.3987  &        0.242     &    -1.650  &         0.099        \\
\textbf{Operation cost}                                          &      -0.7469  &        0.177     &    -4.217  &         0.000        \\
\textbf{Station availability}                                    &       0.3862  &        0.099     &     3.920  &         0.000        \\
\textbf{Sports utility vehicle}                                  &      -0.7066  &        0.838     &    -0.843  &         0.399        \\
\textbf{Sports car}                                              &       0.8799  &        0.852     &     1.032  &         0.302        \\
\textbf{Station wagon}                                           &      -0.1506  &        0.477     &    -0.316  &         0.752        \\
\textbf{Truck}                                                   &      -0.5589  &        0.408     &    -1.370  &         0.171        \\
\textbf{Van}                                                     &      -0.6099  &        0.426     &    -1.433  &         0.152        \\
\textbf{EV}                                                      &      -0.8867  &        0.596     &    -1.489  &         0.137        \\
\textbf{Commute < 5 \& EV}                                       &       0.2088  &        0.084     &     2.490  &         0.013        \\
\textbf{College \& EV}                                           &       0.4586  &        0.110     &     4.175  &         0.000        \\
\textbf{CNG}                                                     &      -0.1985  &        0.605     &    -0.328  &         0.743        \\
\textbf{Methanol}                                                &      -0.5804  &        0.599     &    -0.969  &         0.332        \\
\textbf{College \& Methanol}                                     &       0.2075  &        0.090     &     2.304  &         0.021        \\

\bottomrule
\end{tabular}

\caption{Expanded MNL model (cont'd)}
\label{table:vehicle-choice-expanded-mnl-results-2}
\end{table}

\newpage
\section*{\refname}
\bibliography{checking_draft_v2}

\end{appendices}
\end{document}